\RequirePackage{fix-cm}
\documentclass[twocolumn,epjc3]{svjour3}  
\smartqed  
\RequirePackage{graphicx}
\RequirePackage{amsmath}
\RequirePackage{caption}
\RequirePackage{subcaption}
\RequirePackage{enumitem}
\RequirePackage[english]{babel}
\RequirePackage[autostyle, english = american]{csquotes}
\RequirePackage[switch]{lineno}
\usepackage{comment}
\MakeOuterQuote{"}

\RequirePackage[colorlinks,citecolor=blue,urlcolor=blue,linkcolor=blue]{hyperref}

\newcommand{\Te}{$^{130}$Te }
\newcommand{\Ten}{$^{130}$Te}
\newcommand{\teod}{TeO$_2$ }
\newcommand{\teodn}{TeO$_2$}
\newcommand{\kgy}{kg$\,\cdot\,$yr }
\newcommand{\kgyn}{kg$\,\cdot\,$yr}
\newcommand{\tony}{t$\,\cdot\,$yr }

\newcommand{\onbb}{$0\nu\beta\beta$ }
\newcommand{\onbbn}{$0\nu\beta\beta$}

\journalname{Eur. Phys. J. C}

\begin{document}
\title{End-to-End Data Analysis Methods for the CUORE Experiment}
 
\author{ D.~Q.~Adams\thanksref{USC} \and C.~Alduino\thanksref{USC} \and K.~Alfonso\thanksref{VirginiaTech} \and A.~Armatol\thanksref{LBNLNucSci} \and F.~T.~Avignone~III\thanksref{USC} \and O.~Azzolini\thanksref{INFNLegnaro} \and G.~Bari\thanksref{INFNBologna} \and F.~Bellini\thanksref{Roma,INFNRoma} \and G.~Benato\thanksref{GSSI,LNGS} \and M.~Beretta\thanksref{Milano,INFNMiB}
% ...
}

\thankstext{e1}{e-mail: cuore-spokesperson@lngs.infn.it} 



\institute{Department of Physics and Astronomy, University of South Carolina, Columbia, SC 29208, USA\label{USC}
\and
Center for Neutrino Physics, Virginia Polytechnic Institute and State University, Blacksburg, Virginia 24061,USA\label{VirginiaTech}
\and
Nuclear Science Division, Lawrence Berkeley National Laboratory, Berkeley, CA 94720, USA\label{LBNLNucSci}
\and
INFN -- Laboratori Nazionali di Legnaro, Legnaro (Padova) I-35020, Italy\label{INFNLegnaro}
\and
INFN -- Sezione di Bologna, Bologna I-40127, Italy\label{INFNBologna}
\and
Dipartimento di Fisica, Sapienza Universit\`{a} di Roma, Roma I-00185, Italy\label{Roma}
\and
INFN -- Sezione di Roma, Roma I-00185, Italy\label{INFNRoma}
\and
Gran Sasso Science Institute, L'Aquila I-67100, Italy\label{GSSI}
\and
INFN -- Laboratori Nazionali del Gran Sasso, Assergi (L'Aquila) I-67100, Italy\label{LNGS}
\and
Dipartimento di Fisica, Universit\`{a} di Milano-Bicocca, Milano I-20126, Italy\label{Milano}
\and
INFN -- Sezione di Milano Bicocca, Milano I-20126, Italy\label{INFNMiB}
% ...
}
\thankstext{fn1}{Deceased}
\thankstext{fn2}{Deceased}
\thankstext{fn3}{Presently at: Instituto de Física, Universidade de São Paulo, São Paulo 05508-090, Brazil}

\date{Received: date / Accepted: date}

\maketitle

\begin{abstract}
%
The Cryogenic Underground Observatory for Rare Events (CUORE) experiment set the most stringent limit on the neutrinoless double-beta (\onbbn) decay half-life of $^{130}$Te with 2\,\tony \teodn\ analyzed exposure. In addition to \onbb decay, the CUORE detector---a ton-scale array of nearly 1000 cryogenic calorimeters operating at $\sim$10\,mK---is capable of searching for other rare decays and interactions over a broad energy range. For our searches, we leverage the available information of each calorimeter by performing its optimization, data acquisition, and analysis independently. We describe the analysis tools and methods developed for CUORE and their application to build high-quality datasets for numerous physics searches. In particular, we describe in detail our evaluation of the energy-dependent detector response and signal efficiency used in the most recent search for \onbb decay.  

\textcolor{blue}{Include keywords, PACS and mathematical subject classification numbers as needed}\\

\end{abstract}

\noindent

\section{Introduction} \label{sec:intro}
CUORE (Cryogenic Underground Observatory for Rare Events)~\cite{cuore2025:science,CUORE:2021mvw} is a leading neutrinoless double-beta (\onbbn) decay experiment. 
Compared to standard double-beta decay, in which two neutrons in a nucleus convert into two protons with the emission of two electrons and two anti-neutrinos, only electrons would be emitted in neutrinoless double-beta decay. Given the absence of neutrinos from the final state, the signature for this hypothetical process is a monoenergetic peak at its Q-value (Q$_{\beta\beta}$), which is typically in the few-MeV range. Lower limits on the half-life for this process range between $4.4\,\times\,10^{24}$\,--\,$2\,\times\,10^{26}$\,yr~\cite{cuore2025:science,CUPID:2022puj,Agrawal:2025,GERDA:PRL,KamLAND-Zen:2022tow}. 

The CUORE detector employs cryogenic calorimeters to measure the energy of 
particle interactions and decays. With this technique, the energy deposited by a particle in an absorber material maintained at a temperature of $\sim$10\,mK is converted into thermal phonons (heat) and measured as a temperature variation. The slight 
differences in the intrinsic properties of CUORE detector components---i.e., absorber, temperature sensor, etc.---across calorimeters, and their sheer number, necessitated 
the development of a calorimeter-depen\-dent readout chain~\cite{FrontendCUORE,LinearSupplyCUORE,PulserCUORE} and new automated procedures for the characterization and optimization of various detector parameters~\cite{CUORE:2022hzh,Alfonso:2020yee,Dompe:2020JLTP,DAddabbo:2017efe}. In comparison, predecessor cryogenic calorimetric experiments utilized at most tens of detectors and were on the kilogram scale ~\cite{Alduino:2016vjd,BIASSONI2020103803}.

In our analysis framework, each calorimeter is treat\-ed independently. When a particle deposits energy in a CUORE crystal, a signal pulse forms in the temperature sensor readout, which we measure as voltage. The signal waveform, $v(t)$, of each detector can be modeled as the sum of a detector
response function, $s(t)$, and a noise term, $n(t)$: 
\begin{equation}\label{eq:pulse}
    v(t) = b + A \cdot s(t-t_0) + n(t)
\end{equation}
where $b$ is the DC level (baseline) of the detector sensor, $A$ is the amplitude of the signal, $t$ is time, and $t_0$ is the start time of the pulse relative to a defined event window. 

\begin{figure}
  \includegraphics[width=\columnwidth]{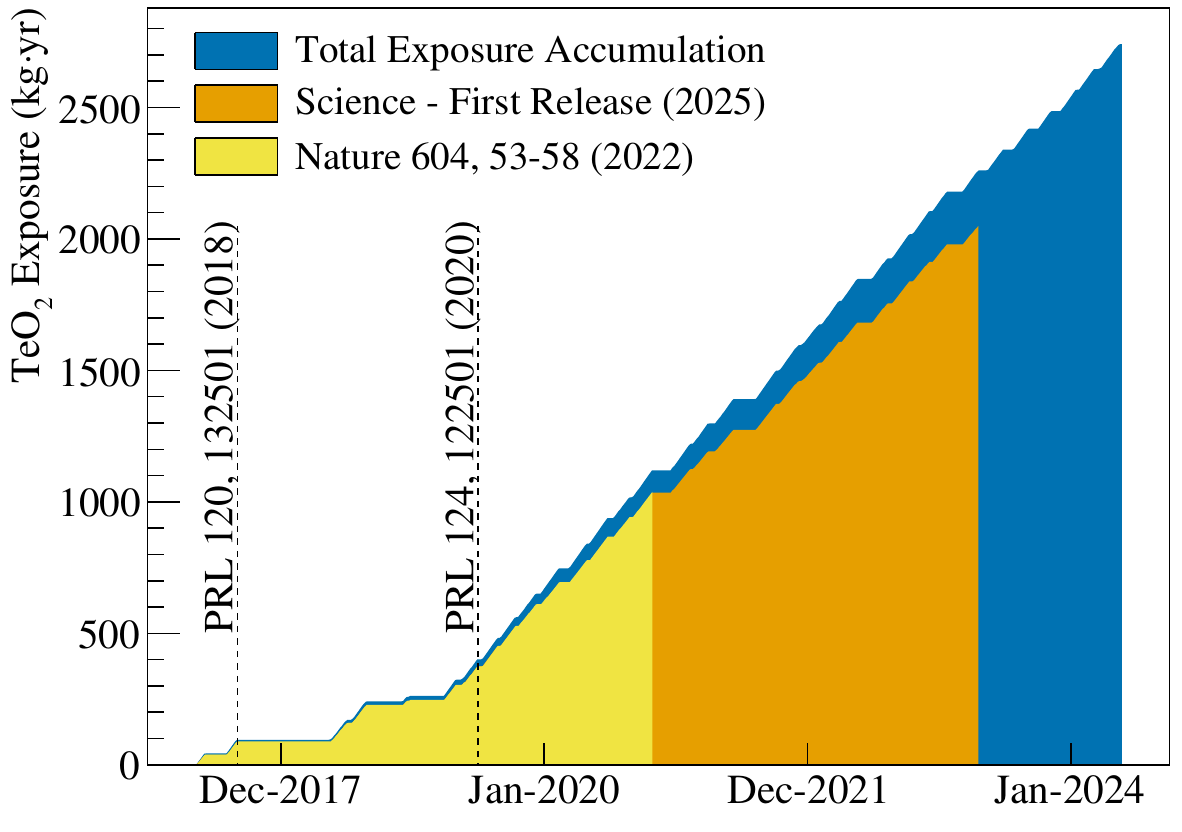}
\caption{Accumulated \teod exposure since the beginning of data-taking is shown in blue. Cryogenic improvements for the operational stability and detectors optimization were implemented between 2017 and 2018~\cite{Nutini:2020vtd,ALDUINO:cryo2019}. From 2019, the experiment has been in stable data taking at 11\,--\,15\,mK, with a strikingly high duty cycle ($\sim$90\%) for the cryogenic calorimetric technology~\cite{CUORE:2021ctv}. \teod exposures after analysis cuts corresponding to major data releases are also shown in yellow and orange. }
\label{fig:exposure2ty}       
\end{figure}

In this paper, we describe how each component in  Eq.~\ref{eq:pulse} is determined for each calorimeter and used to identify and process signal events to build an energy spectrum; evaluate the energy-dependent detector response; and measure related detection efficiencies. Most recently, this set of analysis techniques has been applied to the CUORE 2\,\tony \teod exposure data release (see Fig.~\ref{fig:exposure2ty}).

\section{The CUORE experiment} \label{sec:cuore}

Located at the Gran Sasso National Laboratory in Italy, CUORE primarily searches for \onbb decay of \Ten, whose signature is a peak at 2527.5\,keV~\cite{Rahaman:2011zz}. The CUORE detector consists of 19 towers, each with 13 floors of 2\,$\times$\,2 calorimeter arrays. By using 988 cryogenic calorimeters as particle detectors, CUORE's active detector mass of 742\,kg \teod simultaneously serves as the source of \Te \onbb decay and the means of detection. In each CUORE calorimeter, the absorber material is a TeO$_2$ crystal whose temperature is measured by a neutron-transmutation-doped (NTD) germanium thermistor. Each crystal is also instrumented with a silicon resistor (heater), designed to periodically inject a fixed amount of energy in the detector for thermal gain correction~\cite{ANDREOTTI2012161}. When a CUORE crystal is operated at $\sim$10\,mK, the temperature variation due to energy deposited by a particle is $\sim$100\,$\mu$K/MeV and the corresponding voltage drop across the NTD is $\sim$400\,$\mu$V\!/MeV~\cite{Alfonso:2020yee}. The mean relative energy resolution of the CUORE calorimeters is $\sim$0.3\% at Q$_{\beta\beta}$~\cite{CUORE:2021mvw}.

\subsection{Data taking campaigns} \label{sec:datataking}

Data collection is based on $\sim$24-hour-long runs. We distinguish between physics runs, used to perform rare event searches, and calibration runs, used to calibrate the energy response of our detectors. These runs are combined into datasets. A CUORE dataset contains roughly six weeks of physics runs with a set of calibration runs at the beginning and at the end. Each set of calibration runs typically lasts five days and is generally shared between consecutive datasets. 

\begin{figure}
\includegraphics[trim={5cm 0cm 5cm 1.5cm},clip,width=\columnwidth]{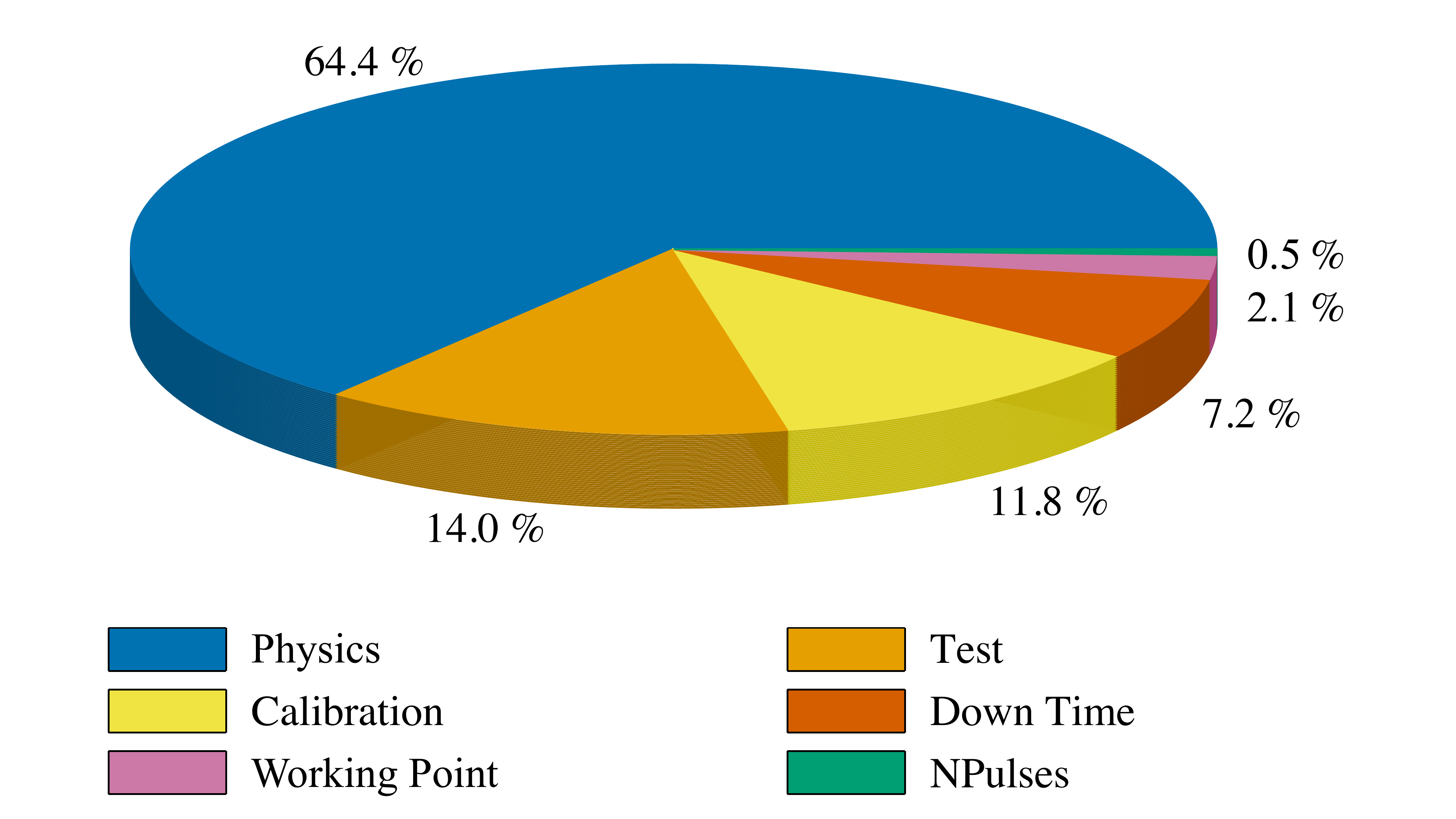}
\caption{Breakdown of the CUORE run-time by measurement type since the beginning of stable data-taking starting in 2019. Down Time corresponds to periods during which no data is recorded.}
\label{fig:piechart}      
\end{figure}

Datasets also include auxiliary measurement types. To monitor the stability of the NTD resistances over time, we perform Working Point (WP) measurements. NPulses measurements, where a scan of the low energy region is performed by injecting heater pulses of variable amplitude, are used to check the stability of the trigger thresholds. Test measurements are acquired during maintenance and optimization operations to monitor for possible changes in data quality.
A pie-chart with the breakdown of the CUORE run-time is shown in Fig.~\ref{fig:piechart}.

The CUORE data releases are summarized in Table \ref{tab:data-releases}. Each release included the re-analysis of the previously available data, since we progressively implemented improvements in the signal triggering and processing algorithms. 
In 2018, between the first two data releases, the operating temperature was lowered from 15.0\,mK to 11.8\,mK to investigate the thermal gain of the detectors. Since mid-2021, the original CUORE operating temperature was restored. The most recent data production includes data from May 2017 to April 2023, for a total of 2\,\tony of analyzed exposure in \teodn~\cite{cuore2025:science,cuore2024:buccinu}. The current data-taking plan is to reach 3\,\tony of analyzed exposure in \teodn, corresponding to $\sim$1\,\tony of analyzed exposure in \Ten, for the \onbb decay search. 
\begin{table}[h] 
\begin{center}
\begin{tabular}{c c c c}
\hline
Date of & Number & Exposure & Ref. \\ 
closing dataset & of datasets & (\kgyn) & \\ 
\hline
Sep 2017 & 2 & 86.3 & \cite{Alduino:2017ehq} \\ 
\hline
Jul 2019 & 7 & 372.5  & \cite{Adams:2019jhp} \\ 
\hline
Dec 2020 & 15 & 1038.4 & \cite{CUORE:2021mvw} \\ 
\hline
Apr 2023 & 28 & 2039.0 & \cite{cuore2025:science} \\ 
\hline
\end{tabular}
\end{center}
\caption{All CUORE data releases start from the first dataset taken in 2017. For each data release, the corresponding date of the closing dataset, number of datasets, analyzed \teod exposure, and publication reference is listed.} 
\label{tab:data-releases}
\end{table}

\subsection{Data acquisition} 
The CUORE Data Acquisition System (DAQ) fulfills different purposes: digitizing analog signals from each of the $\sim$1000 calorimeters, performing the triggering of the data, and storing the detector configuration parameters and data. The acquisition software, \emph{Apollo}~\cite{DiDomizio:2018ldc}, is a custom C++ code based on the ROOT package~\cite{BRUN199781}. 

The voltage stream from each calorimeter is passed through an amplification stage and an anti-aliasing Bes\-sel filter stage. It is digitized with a 1\,kHz sampling frequency by an 18-bit ADC. The DAQ stores the continuous waveforms from all the detectors for offline analysis. For online monitoring, the DAQ also applies an online software trigger (see Sec.~\ref{sec:trigger}) to the data stream for the event construction, where an event is a segment of a waveform around the trigger. The DAQ associates each event with a timestamp and the basic information needed to identify pulses in the data stream. 

\subsection{Data production and analysis}
The CUORE data production consists of a series of digital signal processing algorithms which allow the triggered waveforms to be converted into a calibrated energy spectrum. This is done using the \emph{Diana} software developed by the CUORE collaboration~\cite{Alduino:2016zrl}, a modular C++ custom software, also based on ROOT~\cite{BRUN199781}. By the end of the data production, each triggered thermal pulse is associated with its determined event-based analysis quantities such as energy and shape parameters, and proximity to other triggered events in space and time. Given the large number of detectors in CUORE, manual scrutiny of each analysis step becomes impractical and error-prone. Automated tools have been developed to systematically and efficiently cross-check the data processing.

\section{Reducing the noise in CUORE}

The signal bandwidth of the CUORE detector output is 0\,--\,20\,Hz. The dominant noise in this bandwith comes from 4 Pulse Tube cryocoolers (PTs), which cool down and maintain the 40\,K and 4\,K stages of the cryostat~\cite{ALDUINO:cryo2019}. These PTs introduce vibrational noise between 0.1\,--\,1.5\,Hz due to their rotating motor.  In particular, a prominent 1.4-Hz (and harmonics) frequency is generated by the speed of the rotating motor. To reduce the impact of these vibrations, a system to control individual PT pressure oscillations was developed~\cite{DAddabbo:2017efe}. Between datasets, the relative phases of the PTs are optimized and tuned to the configuration that minimizes their induced noise.

Software denoising techniques are used to further suppress PT and seismic noise in our detector measurements. Our denoising algorithm correlates the noise measured by each calorimeter with the noise measured by auxiliary devices~\cite{vetter2024improving}. These correlations are used to predict and subtract vibrationally-induced noise from the continuous CUORE data stream. The auxiliary devices, which include accelerometers, seismometers, and microphones, are sensitive to external vibrational and acoustic noise sources in the range of 0.1\,Hz\,--\,1\,kHz; they are installed in various locations over the cryostat in the CUORE Faraday room~\cite{Bucci_2017} and their measurements of the environmental noise are digitized with the same system used for the CUORE detectors. 

As part of the algorithm, we create a vector of cross-spectral densities with each auxiliary device for each calorimeter. The spectral density of the calorimeter is determined using only its noise events (waveforms without pulses). The auxiliary device terms in the vector include both linear and quadratic signal contributions in the time domain; the former captures the capacitive pickup from vibrations in the cabling and the latter is a proxy for the unipolar heat response due to the vibrations in the calorimeters. The product of this vector and the inverse of the matrix of cross-spectral densities of all auxiliary device pairs results in a transfer function from auxiliary device output to the calorimeter noise at a particular frequency. The transfer function for each auxiliary device is convolved with the continuous auxiliary device waveform. The sum of the resulting convolutions over all auxiliary devices is the predicted calorimeter noise. The result of subtracting this predicted noise from the original continuous data in the time domain is the \emph{denoised} data. 

The denoising algorithm is applied to each calorimeter independently for every physics run; this accounts for daily fluctuations in the noise. Calibration runs, however, last closer to $\sim$30 hours and have event rates 10\,--\,100 times higher than physics runs, resulting in too few noise events to construct reliable transfer functions from the calibration data alone. In this case, we average transfer functions from $\sim$1 week of physics runs taken immediately before or after the calibration period. While this averaging attenuates run-level noise features, it makes the denoising of calibration data possible. 

\begin{figure}
  \centering
  \includegraphics[trim={0.4cm 0cm 1.75cm 1cm},clip,width=\columnwidth]{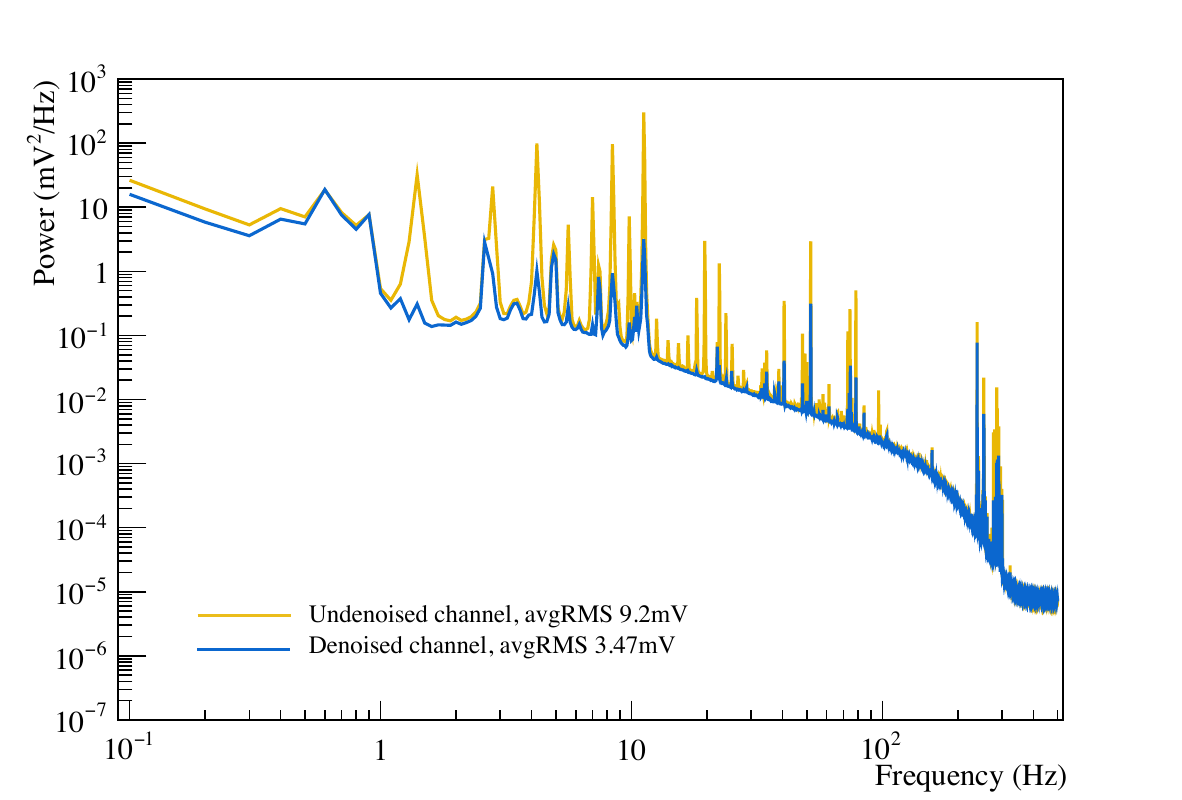}
\caption{Noise spectral shape of a single channel-dataset (Ch-DS) before and after the denoising using accelerometers and seismometers.}
\label{fig:denoising}       
\end{figure}

Acting as a time-dependent filter, this method removes steady-state noise (e.g., PT noise) as well as transient noise, such as anthropogenic activity near the detector. The reduction of the peak amplitudes, particularly at $\sim$1.4\,Hz and its harmonics, in the denoised power spectrum (blue) compared to the undenoised po\-wer spectrum (orange) in Fig.~\ref{fig:denoising} demonstrates the effect of this software technique.

\section{Data trigger}\label{sec:trigger} 
\begin{figure}
    \centering
  \includegraphics[trim={0cm 1.8cm 0cm 0.5cm},clip,width=\columnwidth]{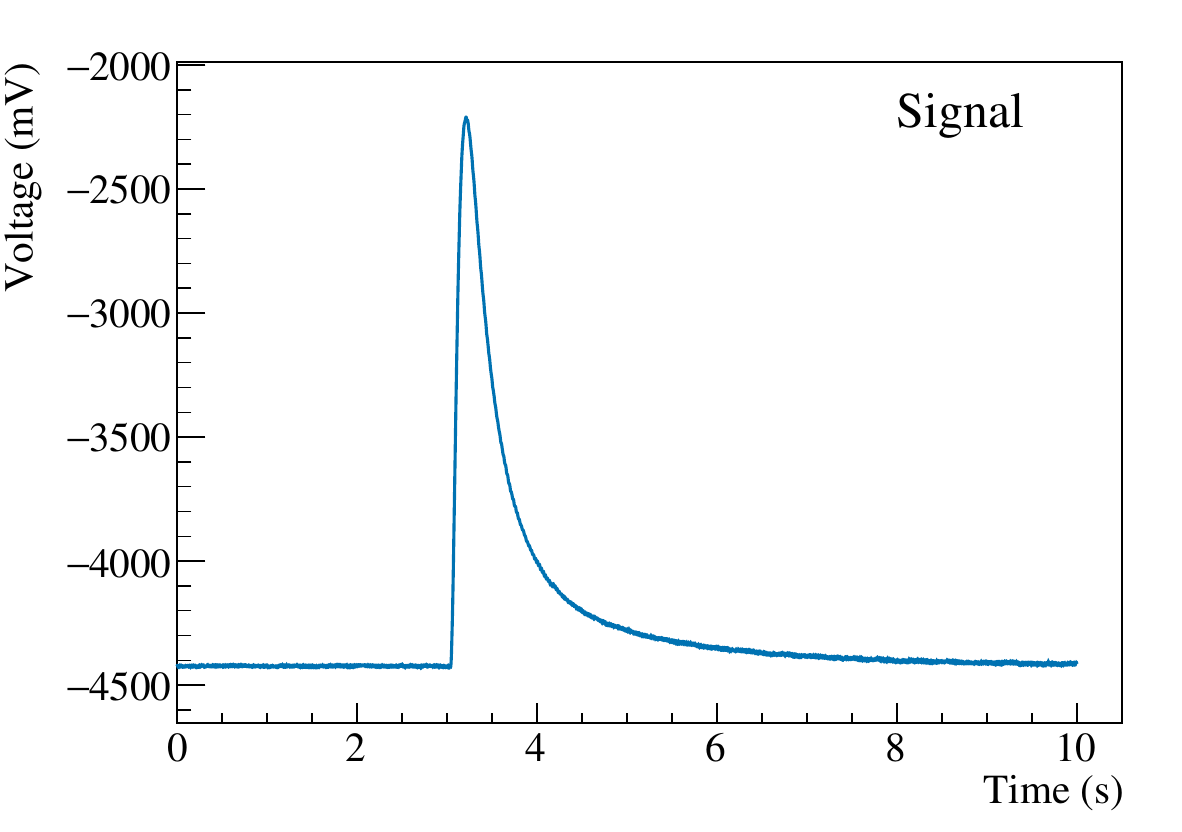}
  \includegraphics[trim={0cm 1.8cm 0cm 0.5cm},clip,width=\columnwidth]{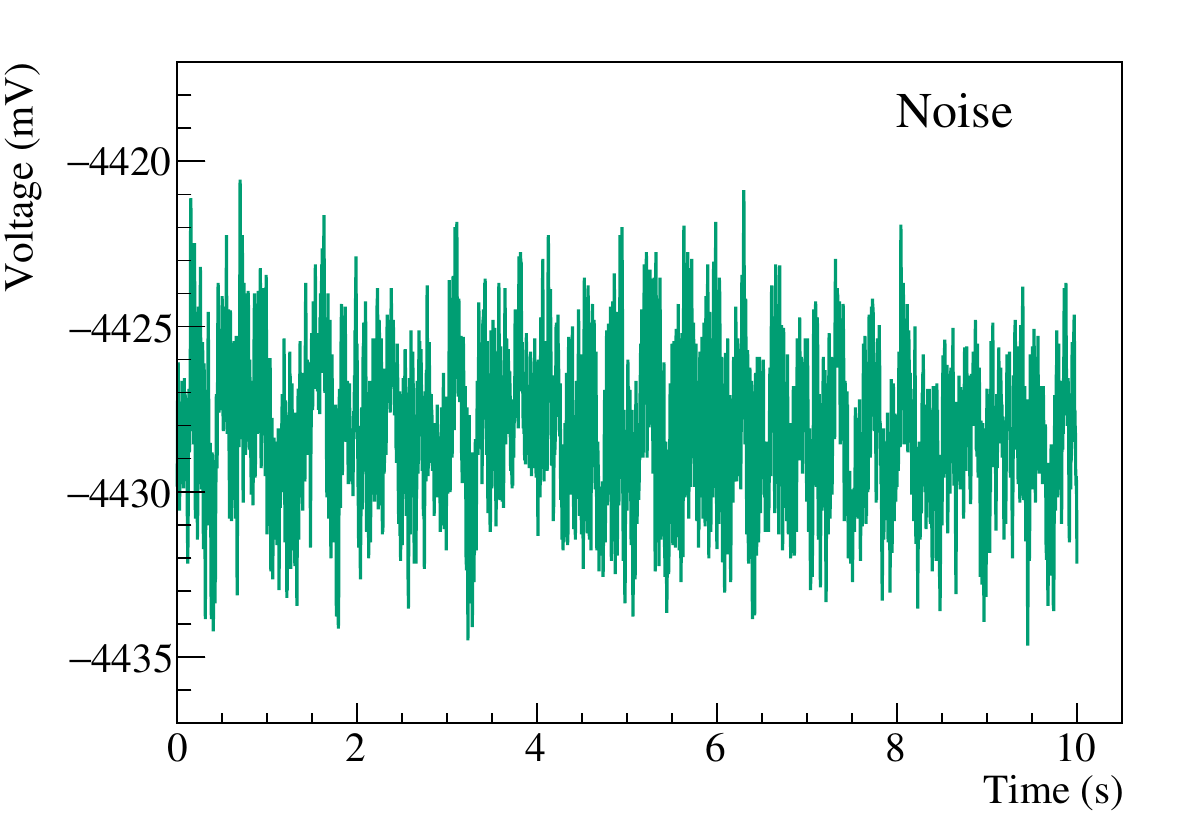}
  \includegraphics[trim={0cm 0cm 0cm 0.5cm},clip,width=\columnwidth]{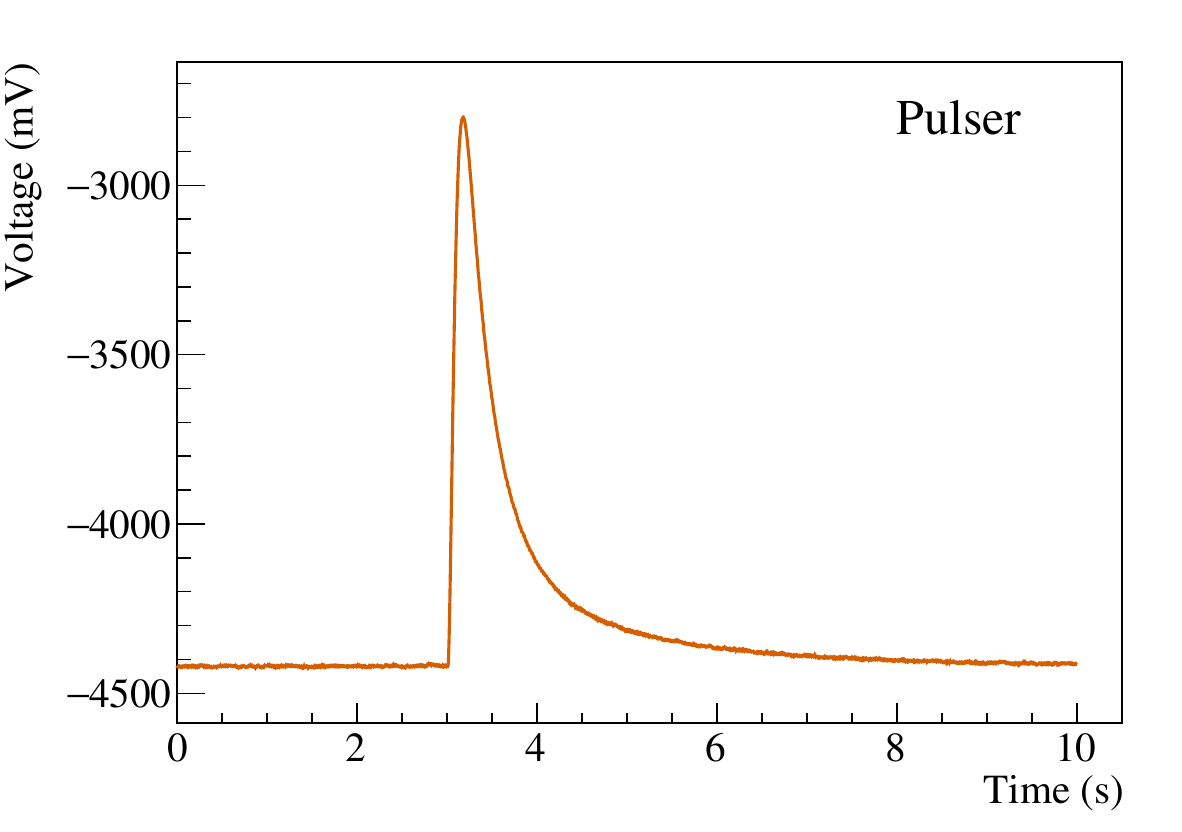}
\caption{Events flagged by the three trigger types: signal, noise, and pulser. In general, the signal waveforms exhibit rise times (from 10\,--\,90\% of the pulse height) of 50\,--\,200\,ms and decay times (from 90\,--\,30\% of the pulse height) of 500\,ms\,--\,1\,s. 
}
\label{fig:triggers}       
\end{figure}

In CUORE, three different types of trigger flags are implemented: signal, noise, and pulser~\cite{DiDomizio:2018ldc}. Signal triggers identify particle-induced pulses, noise triggers identify potential segments of the data to use for noise studies, and pulser triggers identify silicon heater-induced pulses. Each trigger is associated with a 10-s event window, which begins 3\,s before the trigger. All three trigger algorithms run in parallel during the data acquisition. An example of an event window, for each trigger type, is shown in Fig.~\ref{fig:triggers}.

Two types of software triggers are used to identify signal events, one runs online during the data-taking, the other runs offline on the continuous data; both employ calorimeter-dependent thresholds. Typical signal trigger rates in CUORE are on the order of 1\,--\,10\,mHz during physics runs, and 50\,--\,250\,mHz during calibration runs. To monitor the detector stability and characterize the noise, we force the trigger of random events every 80\,s. In most cases, the corresponding triggered events will contain no physical pulse. Thus, after excluding event windows coincidentally containing a signal and/or pulser trigger, the random events can be used to evaluate the noise (see Sec.~\ref{of}). To generate fixed thermal energy events for thermal gain studies (see Sec.~\ref{hTGS}), we apply a voltage across the silicon resistors on the crystal to inject a precise amount of heat.  
These events and are flagged with a pulser trigger.

\subsection{Online derivative trigger}

CUORE uses an online trigger to identify particle-induced pulses over a range of energies during the data-taking. The purpose of this trigger is to enable low-latency monitoring of the detector performances and provide a priori signal and noise information needed by the more sophisticated offline trigger algorithm (see Sec.~\ref{sec:OT}). The online trigger is integrated in the $Apollo$ software. The algorithm is a derivative trigger (DT) which fires whenever the derivative of the waveform, calculated on a given time interval (\emph{average}), exceeds a threshold for a specified \emph{debounce} time. Both average and debounce intervals are set to 40\,ms~\cite{DiDomizio:2018ldc}. An artificial trigger dead time---i.e., minimum time interval between consecutive triggers in the same calorimeter---is set to 200\,ms. 

To identify low energy events, the threshold for each calorimeter is set just above the noise level. Since the noise varies among the calorimeters, the CUORE thresholds are also calorimeter dependent. 
We set the trigger threshold based on the trigger rate on noise events. For each calorimeter, we first acquire dedicated test runs where no heat pulses are injected and then apply an offline algorithm that simulates the online Apollo derivative trigger using different thresholds. After computing the associated trigger rates, the optimal threshold is identified as the lowest threshold for which the trigger rate simultaneously is lower than a few mHz and demonstrates a shallow dependence on the threshold. The first condition ensures that the DAQ is not overwhelmed by the incoming trigger rate, while the second condition mitigates triggers on noise fluctuations.

\subsection{Offline Optimum Trigger}\label{sec:OT}

The availability of continuous data allows us to digitally re-trigger it at a later time, using different trigger settings and different triggering algorithms. For the official data production, we use the Optimum Trigger (OT) algorithm~\cite{Alduino:2017xpk,Branca:2020bra}. The OT is based on the matched filter technique and uses the expected signal pulse shape from DT-identified pulses in the denoised data to enable significantly lower thresholds with respect to the online DT. The buffer of the data stream is filtered with an Optimum Filter (OF) (see Sec.~\ref{of}) in the frequency domain, in order to maximize the signal-to-noise ratio. Since the signal has a relatively small bandwidth ($<$30\,Hz), the acquired data are initially passed through a low-pass Chebyshev filter using a cutoff value of 5\% of the sampling frequency, and then down-sampled from 1\,kHz to 125\,Hz. This is done to  speed up the algorithm application while keeping the same performance within the signal bandwidth. Thus, each buffer is filtered with the OF transfer function, then transformed back to the time domain, where a trigger threshold is set to 4 times the noise resolution. The filtered waveforms are less noisy than the original ones and the baseline fluctuations are reduced. Furthermore, because the filter is sensitive to the shape of the expected signal, triggers on spurious noise-induced pulses are suppressed.

\section{Thermal pulse amplitude reconstruction} 

The energy deposited to a calorimeter is correlated to the resulting pulse amplitude. Thus, amplitude reconstruction is the precursor to building the energy spectrum for our rare event searches. As a first order approximation, the pulse amplitude ($A$ in Eq.~\ref{eq:pulse}), can be decomposed in terms of an energy-dependent term $a(E)$, where $E$ is the energy deposited into the crystal and a temperature-dependent term $G(T)$, which represents the detector intrinsic gain related to the operating temperature $T$: 
\begin{equation}\label{eq:A}
    A(E,T) \approx a(E) \cdot G(T)
\end{equation}
Thus, we aim to extract $a$ from our measured value of $A$ for each pulse. 
In order to accurately evaluate the energy-dependent amplitude component of each event, we first assess the quality of each event pulse (Sec.~\ref{prepro}), then construct an OF from a subset of high quality events and apply the filter to each event (Sec.~\ref{of}), finally we characterize and remove the temperature dependence from the measured pulse amplitudes (Sec.~\ref{stab}).

\subsection{Thermal pulse basic parameters} \label{prepro}

\begin{figure}
    \centering
  \includegraphics[trim={0cm 1.8cm 0cm 0.5cm},clip,width=\columnwidth]{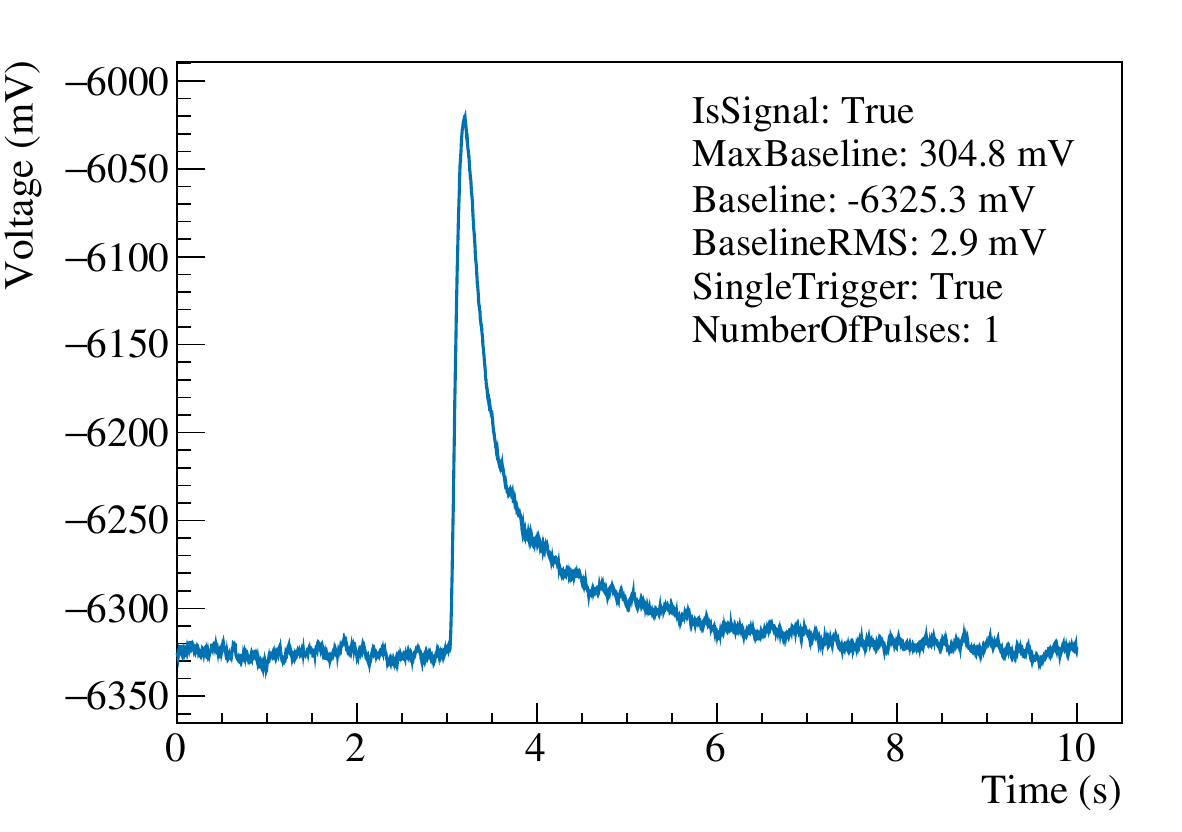}
  \includegraphics[trim={0cm 0cm 0cm 0.5cm},clip,width=\columnwidth]{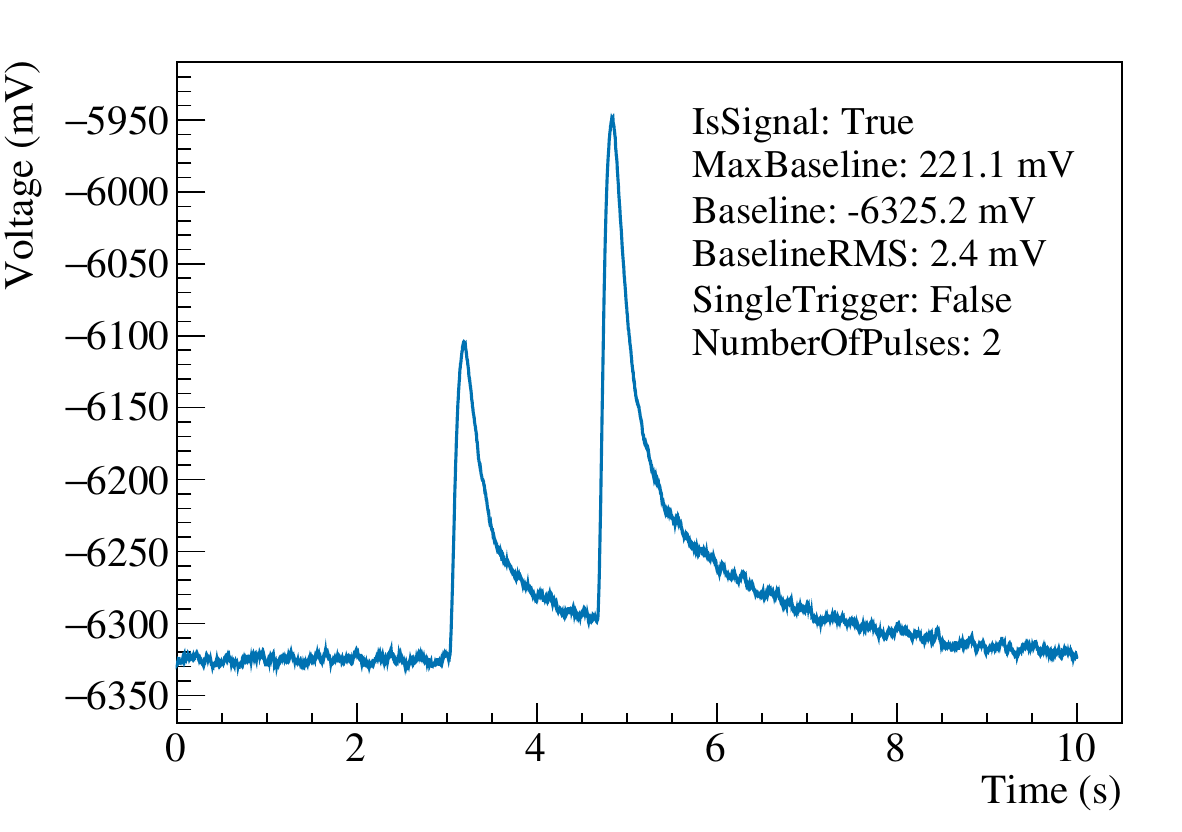}
\caption{Example of single pulse and pile-up events. The variables associated with the main event are listed. The 10-s event window begins $\sim$3\,s before the main event.} 
\label{fig:pulses}       
\end{figure}

We perform a preliminary processing of the denoised data to evaluate basic parameters of the event. The pretrigger, defined as the first 2.25\,s of each event window, is used to compute the baseline parameters which identify the detector conditions before the particle interaction in the detector. The $Baseline$ is calculated as the average of the pretrigger and is used as a proxy for the calorimeter temperature at the time of the event. The pretrigger is fit with a first-order polynomial, and the slope ($BaselineSlope$) is used to identify events overlapping with the decaying tail of an earlier event in the same calorimeter. A proxy for the raw noise is given by the $BaselineRMS$, which is calculated as the root mean square error of the linear fit used to compute the baseline slope. Pile up events are identified by the $SingleTrigger$ and $NumberOfPulses$ parameters, which track the number of triggers and signal pulses, respectively, within the entire 10-s window. Finally, a simple estimate of the pulse amplitude ($MaxBaseline$) is obtained as the raw pulse height relative to the $Baseline$. Examples of CUORE pulses and associated basic parameters are reported in Fig.~\ref{fig:pulses}.

\subsection{Optimum Filter (OF)} \label{of}
The OF technique~\cite{Gatti:1986cw} is used to obtain a more precise estimator for the pulse amplitude. The OF is designed to maximize the signal-to-noise ratio given a known signal shape and noise power spectrum. The detector response function $s(t)$ is assumed to be a fixed template. 
Moreover, the stochastic noise on the detector is assumed to have a stationary behavior in order to be described by a definite noise function $n(t)$. Thus, the OF pulse in the frequency domain can be written as
\begin{equation}
    V^{OF}(\omega) \propto \frac{S(\omega)^*}{|N(\omega)|^2} e^{i \omega \tau_M} V(\omega)
\end{equation}
where $V(\omega)$ and $S(\omega)$ are the Fourier transforms of the signal and the detector response function, respectively; $|N(\omega)|^2$ is the noise power spectral density; $\omega$ is the frequency; and $\tau_M$ acts as a phase shift which maximizes the amplitude of the filtered signal transformed back to the time domain. 

\begin{figure}
  \includegraphics[trim={0cm 0cm 2.cm 1.1cm},clip,width=\columnwidth]{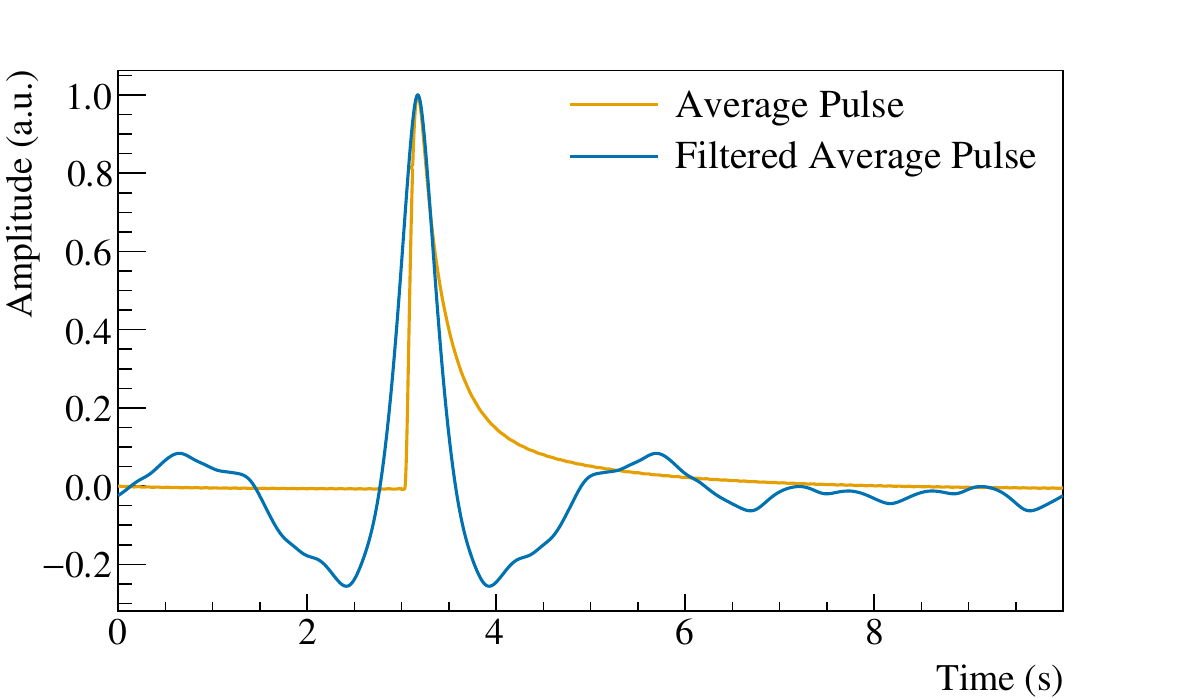}
\caption{Example of an average pulse (orange) for a single Ch-DS. The filtered average pulse (blue) demonstrates the effect of the OF on the average pulse.} 
\label{fig:AP}       
\end{figure}

\begin{figure}
  \includegraphics[trim={0cm 0cm 2.cm 0.9cm},clip,width=\columnwidth]{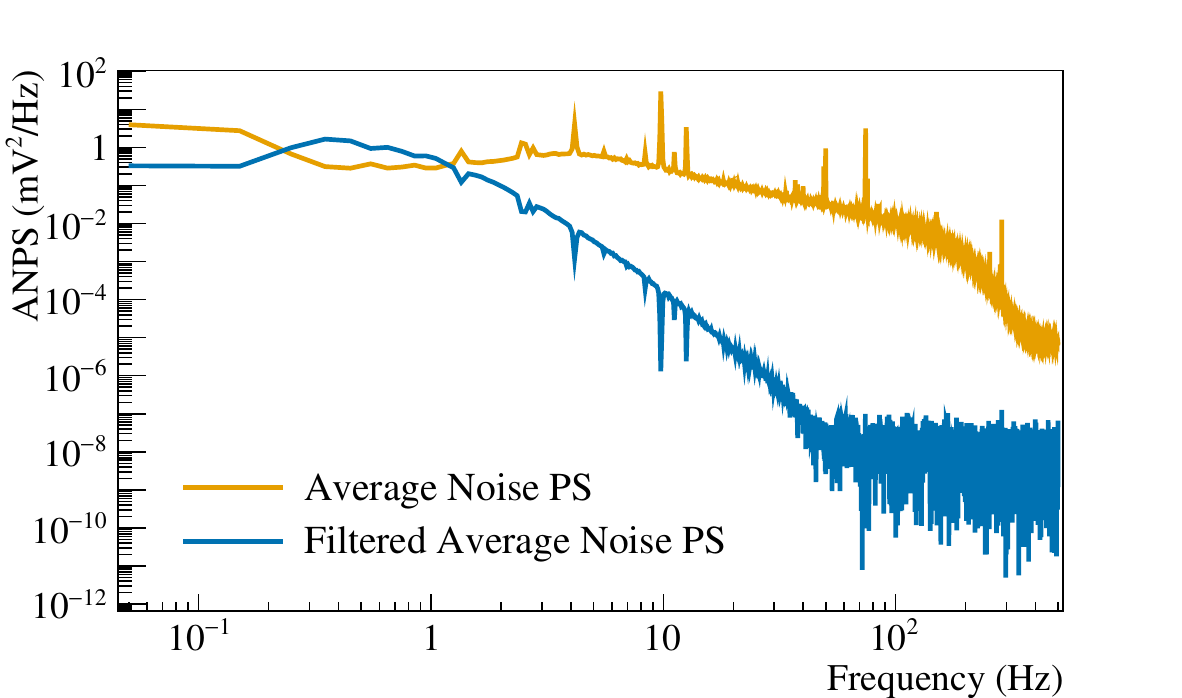}
\caption{Example of an average noise power spectrum (orange) for a single Ch-DS. The filtered average noise power spectrum (blue) demonstrates the effect of the OF on the frequency components.} 
\label{fig:ANPS}       
\end{figure}

Since each calorimeter $i$ has different characteristics reflected in the pulse shape and noise, both the response function $s_i(t)$ and the noise power spectrum $|N_i(\omega)|^2$ must be evaluated for each detector. We use proxies for building the filter templates: a calorimeter-dependent averaged pulse (AP) for $s_i(t)$ and an averaged noise power spectrum (ANPS) for $|N_i(\omega)|^2$. Based on the first-level analysis (see Sec.~\ref{prepro}), we apply relevant selection cuts on the pulse parameters to identify genuine particle signals which are used to build the AP. In particular, the APs are built from calibration data given the larger available statistics of signal pulses across multiple energies. A dedicated selection is also applied to the noise triggers to identify pure noise events (waveforms in which no pulses occurs) to build the ANPS. Table~\ref{tab:ap-anps} summarizes the selections for building the AP and ANPS. Examples of a built AP and ANPS are shown in Fig.~\ref{fig:AP} and Fig.~\ref{fig:ANPS}, respectively. 

After we construct the calorimeter-dependent filters, we apply the corresponding filter to each event. The amplitude of each filtered pulse is determined by interpolating the data points around the maximum with a second-order polynomial in order to reduce the discretization error of the evaluation.

\begin{table}[] 
\begin{tabular}{l  l  l}
\hline
 & AP & ANPS \\ 
\hline
Event Type & Signal & Noise \\ 
SingleTrigger & True & True \\
NumberOfPulses & 1 & 0 \\
$|$BaselineSlope$|$ & \textless{}0.005\,mV/s & \textless{}0.001\,mV/s \\ 
Raw amplitude & \textgreater40$\cdot$BaselineRMS & \textless10$\cdot$BaselineRMS \\ 
Data Type & Calibration & Background \\ 
\hline
\end{tabular}
\caption{List of the event selections for building AP and ANPS. The Baseline Slope cut for the ANPS is set over the whole event window, while AP only uses the pretrigger. 
\label{tab:ap-anps}}
\end{table}

\subsection{Thermal gain stabilization} \label{stab}  

Small variations of the base temperature may change the calorimeter intrinsic gain, leading to different pulse amplitudes for events depositing the same energy in a given crystal. This degrades the detector energy resolution. Therefore, it is necessary to correct the filtered pulse amplitude against thermal drifts that occur during the data taking. Events depositing the same energy can be used to trace the evolution of the internal gain as a function of the crystal temperature. 
A proxy for the temperature is the pulse baseline:   
\begin{equation}\label{eq:bsl}
    b(T) = V_{\mathrm{off}} - G_{\mathrm{ele}} \cdot V_{\mathrm{out}}(T)
\end{equation} 
where $V_{\mathrm{off}}$ and $G_{\mathrm{ele}}$ are respectively the offset and gain from the electronics, 
and $V_{\mathrm{out}}$ is the NTD output voltage. In general, a decrease of the baseline corresponds to a cooling of the detector, which results in a slightly higher pulse amplitude since the detector internal gain increases.

CUORE applies two different techniques for the thermal gain stabilization (TGS): one is heater-based (\emph{hea\-ter-TGS}) and the other is calibration-based (\emph{calibration-TGS})~\cite{Alduino:2016zrl,ANDREOTTI2011822,Alessandrello:1998}. Both approaches are calorimeter dependent. Calibration-TGS was originally developed for calorimeters with faulty or malfunctioning heaters, but was also found to improve the performance of a subset of calorimeters. These calorimeters tend to exhibit different trends in their pulse amplitude with temperature between heater- and particle-induce events. Thus, when available, both approaches are applied to each calorimeter and their relative performance is used for the final estimator selection (see Sec.~\ref{sec:esel}).

\subsubsection{Heater-based stabilization} \label{hTGS}
\begin{figure}
  \includegraphics[trim={0cm 0cm 0.cm 0.9cm},clip,width=\columnwidth]{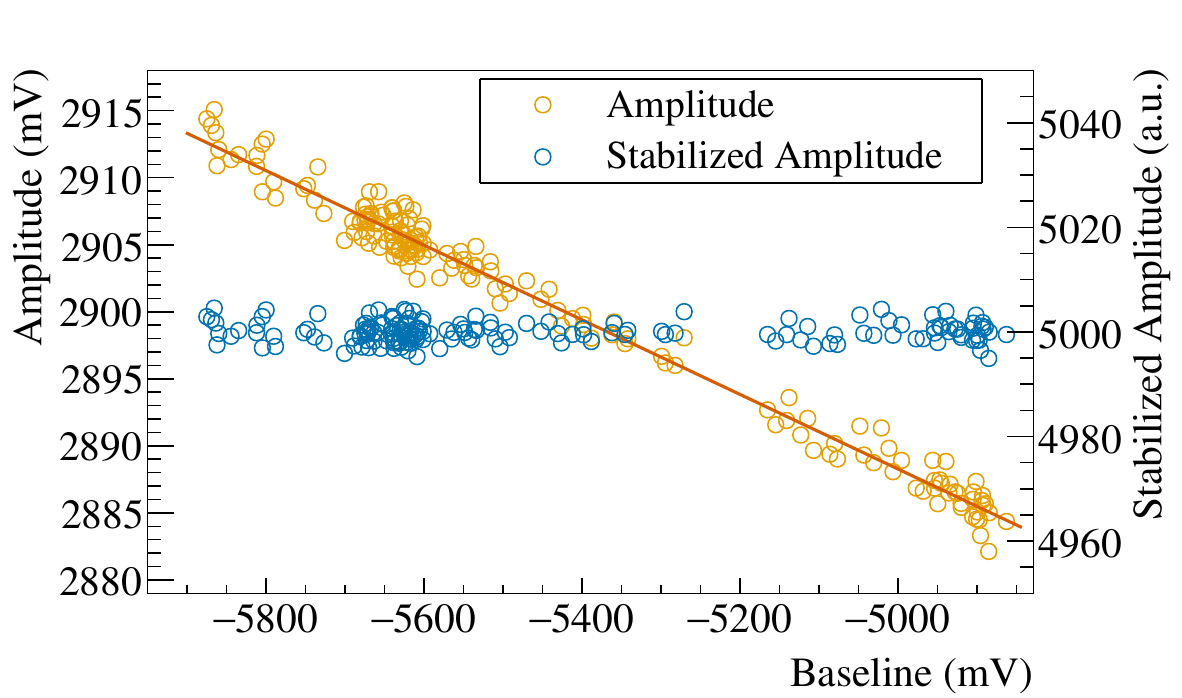}
\caption{Amplitude of pulser events from two runs for a single channel before (orange) and after (blue) thermal gain stabilization. The stabilized amplitudes of the pulser events are set to an arbitrary values of 5000 a.u.}
\label{fig:stab}       
\end{figure}

The heater-based stabilization algorithm relies on the constant voltage signals that are injected into the Si heaters attached to the crystals~\cite{PulserCUORE}. The amplitude of this \emph{stabilization pulser} is selected such that their reconstructed energies represent events at Q$_{\beta\beta}$ ($\sim$3\,MeV). The temperature dependence of the fixed-energy heater pulse amplitudes is determined by using the corresponding $Baseline$, $b$, as a proxy for temperature, $T$. The trend of the temperature-dependent term in Eq.~{\ref{eq:A}}, $G(T)$, is parametrized with a first-order polynomial, so the heater pulse amplitude can be written as
\begin{equation}\label{eq:stab2}
    A_h(T) = a_h(E_h) \cdot (p_0 + p_1 \cdot b(T)), 
\end{equation}
where $a_h$ is a constant (since $E_h$ is fixed) that is set to an arbitrary value and $p_0$ and $p_1$ are the fit parameters. Since $G(T)$ is expected to be the same for both signal and heater pulses, we apply the following correction to each event to extract the energy-dependent amplitude of the signal pulse: 
\begin{equation}\label{eq:stab1}
    a_s(E) = \frac{A(E,T)}{p_0 + p_1 \cdot b(T)}
\end{equation}
where $A$ is the OF-amplitude of the event. This stabilization procedure is performed for each run. Figure~\ref{fig:stab} shows the original and stabilized amplitude against baseline for stabilization pulser events for a single calorimeter.

\subsubsection{Calibration-based stabilization}

The calibration-based stabilization algorithm uses the events in the $^{208}$Tl 2615\,keV gamma peak from calibration data as monoenergetic reference events. Given the limited statistics of the reference pulses, this procedure cannot be applied to a single run, but instead must be applied to the entire dataset. This stabilization algorithm first uses the value of $V_{\mathrm{off}}$---computed for each calorimeter from the WP measurement (see Sec.~\ref{sec:datataking}) that is measured closest in time---to correct the baseline of each reference event. Then, for each calorimeter, the amplitude of the 2615\,keV events as function of $(G_{\mathrm{ele}}\,\cdot\,V_{\mathrm{out}})$ is fit with a quadratic function. Finally, the pulse amplitudes in the dataset are corrected using the corresponding calorimeter-dependent function. Given the potentially larger range of baseline variation between initial and final calibrations, compared to that of a single run, this approach accounts for nonlinearities in the gain temperature dependence.

\section{Energy reconstruction} 
To reconstruct the energy associated with each pulse, we periodically expose the detector array to gamma rays of known energies by suspending $^{232}$Th and $^{60}$Co radioactive sources between the outermost cryostat vessel and external lead shield. For each calorimeter, the amplitude response from each stabilization is calibrated against the characteristic energies. To select between the energy estimators derived from the heater-based and calibration-based stabilizations for each calorimeter, we compare their $^{208}$Tl 2615\,keV peaks.

\subsection{Energy calibration} \label{EnergyCalibration}
\begin{figure}
  \includegraphics[trim={0.1cm 0cm 1.3cm 0.9cm},clip,width=\columnwidth]{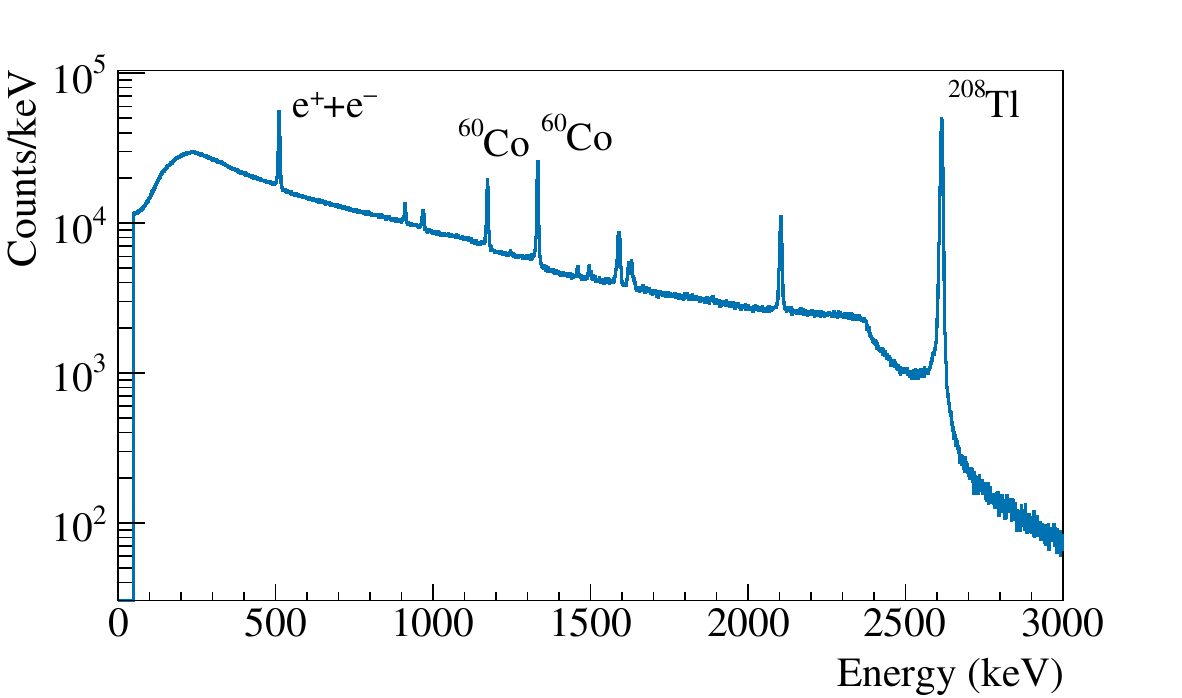}
\caption{Cumulative Th-Co energy calibrated spectrum for a single dataset using the heater-based thermal gain stabilization technique. The peaks used to calibrate the spectrum are at 510.999\,keV (e$^+$\!+e$^-$), 1173.240\,keV ($^{60}$Co), 1332.508\,keV ($^{60}$Co), and 2614.511\,keV ($^{208}$Tl). The other peaks in the spectrum are not used for calibration due to their low statistics in individual detectors.}
\label{fig:calibSpectra}      
\end{figure}

To convert the stabilized pulse amplitudes to energies, we histogram the amplitudes and fit the position of the peaks in the spectrum. Depending on the acquired statistics, we fit 2\,--\,4 peaks for each calorimeter. Among the peaks in the spectrum, the algorithm identifies the highest amplitude peak with the best signal-to-noise ratio as the reference peak. The other calibration peaks are identified by considering their relative positions to the reference peak. We fit two models to each peak. The first model consists of a Gaussian function, while the second model consists of a Crystal Ball function; in both models, we overlap a background contribution---the sum of a first-order polynomial and an error function (Compton edge component). To facilitate the fit of the latter, the Gaussian fit parameter results from the former are used to seed the parameters (mean and width) of the Crystal Ball. We compare the results of both fits and then choose the better performing one for each peak based on convergence and goodness-of-fit. After selecting the model for each peak, we correlate their positions against the calibration gamma-ray energies with a second-order polynomial with zero intercept. All calibration peaks, including the reference peak corresponding to the $^{208}$Tl gamma line at 2615\,keV, are labeled for a representative CUORE spectrum in Fig.~\ref{fig:calibSpectra}. This calibration procedure is performed for each stabilization method available for each calorimeter and dataset. 

To guarantee the fit performance of most calorimeters, we flag calorimeters for manual seeding of the peak positions if any of the following conditions apply: 
\begin{itemize}
    \item the number of events in the calibration spectrum does not meet a set threshold
    \item the algorithm fails to identify the reference peak
    \item the algorithm fails to identify or fit one or more of the non-reference peaks 
    \item the algorithm fails to fit both models to the reference peak 
    \item the fit for the calibration function fails
\end{itemize}
For these calorimeters, we then repeat the calibration(s). 

\subsection{Energy-estimator selection} \label{sec:esel}
When available, we select between the two energy estimators by calculating a figure-of-merit (FOM) for each and taking their ratio: 
\begin{equation} \label{eq:W}
     W \propto \left ( \frac{C_h^2}{\sigma_h^2} \right ) \bigg/ \left ( \frac{C_c^2}{\sigma_c^2} \right ).
\end{equation}
The numerator and denominator in Eq.~\ref{eq:W} correspond to the heater-based ($h$) FOM and the calibration-based ($c$) FOM, respectively, where $C$ and $\sigma$ are the fitted integral and standard deviation of the 2615\,keV peak. We evaluate the Fisher test statistic, $F(\alpha)$ for ($C-1$) degrees of freedom, where $\alpha$ is the significance level---i.e., the probability that an F-value will be larger than the 
$F(\alpha)$. When $W < 1$ and $ W > F(\alpha)$, the calibration-based energy estimator is selected, otherwise the heater-based energy estimator is selected. For a typical dataset, when using $\alpha$\,=\,0.20, this method selects the calibration-based energy estimator for $\sim$100 calorimeters. This value of $\alpha$ was chosen based on the performance of a subset of datasets.

\subsection{Energy thresholds} \label{thresholds}

To evaluate the trigger threshold and the trigger efficiency curve, we use a software procedure to generate and inject simulated pulses of various energies into the data stream. For each CUORE run and each calorimeter, we revert the injected pulse energy to a stabilized pulse amplitude and eventually to a raw pulse amplitude using the calibration (from Sec.~\ref{EnergyCalibration}) and stabilization (from Sec.~\ref{stab}) coefficients calculated in previous steps of the data analysis. We then scale the corresponding average pulse to match the raw amplitude value and overlap it with a random noise waveform, characteristic of the given calorimeter for the dataset under analysis. This emulates a real pulse corresponding to the desired energy. For each calorimeter, we vary the energy of the injected pulses and estimate the trigger efficiency as the ratio of the number of OT triggered events (Sec.~\ref{of}) to the number of injected ones. The event is considered if an OT trigger is found within a time interval corresponding to the width at half maximum of its AP. We fit the efficiency of the OT trigger as a function of the pulse energy using an error function. Then we invert the relation to determine the energy threshold as the energy at which the efficiency function reaches 90\%~\cite{cuore2025:lowE}.

\subsection{Recovery of saturated pulses} 

The CUORE analog electronics boards are saturated for detector voltages $|V_{\mathrm{out}}|>9.2$\,V, corresponding to energies above $\sim$15\,MeV~\cite{CUORE:2024rbd}. To reconstruct the energy of these saturated events for physics analyses at high energies---e.g., the study of the muon spectrum, the search of events induced by neutron captures, or the search for tri-nucleon decay---we utilize a ``time-above-threshold'' metric. With this method, we first consider unsaturated pulses with known energies and for each, measure the time interval ($t_{\mathrm{above}}$) that the pulse exceeds a given threshold voltage ($V_{\mathrm{thresh}}$). We then use the relation between $t_{\mathrm{above}}$ and $h/E$, where $h=V_{\mathrm{thresh}} - Baseline$ is the pulse height and $E$ is the energy of the pulse, to determine the energy of the saturated pulse. Assuming pulse linearity, for a pulse saturating at $V_\mathrm{sat}$, we extrapolate its energy to 
\begin{equation}  \label{eq:Esat} 
    E_\mathrm{sat} = \frac{V_{\mathrm{sat}} - Baseline_\mathrm{sat}}{\frac{h}{E}(t_\mathrm{above}=t_\mathrm{sat})}
\end{equation}
where the denominator is $h/E$ evaluated at $t_\mathrm{sat}$, the time interval above the saturation threshold, $V_{\mathrm{sat}}$.

\section{Reconstruction of multi-site events}\label{MultiSite}
From our Monte Carlo simulations of CUORE, the probability for full absorption of a hypothetical \onbb decay in the same crystal of origin is $\sim$88\%~\cite{Alduino_2017}. Thus, we can leverage the single-site nature of \onbb decay to mitigate the background 
by using an anticoincidence cut to filter away events that occur within a given time interval of another event belonging to a different calorimeter in the detector array. The time interval for the cut, called the coincidence window, is chosen to be wider than the characteristic time resolution of the CUORE calorimeters in order to efficiently identify multi-site events, but also narrow enough to reduce the rate of random coincidences. This cut primarily rejects $\alpha$ decays that occur on the crystal surfaces, $\gamma$ rays that scatter in one calorimeter before being absorbed in another, cascade $\gamma$ rays from radioactive decays, and muons passing through the tower(s) along with their secondary neutrons. Thus, we categorize events by their multiplicity, M, defined as the number of triggered pulses within the coincidence window, $\Delta t^*$.

\subsection{Time response synchronization} \label{TimeSync}
A true physical coincidence among different calorimeters originates from a particle passing through and depositing energy in more than one crystal, or from multiple particles emitted simultaneously from the same radioactive decay depositing energy in different crystals. In both cases, the time difference ($\Delta t$) between the energy depositions is negligible with respect to the intrinsic pulse rise time of the calorimeter. When determining M, we specify both the coincidence window, $\Delta t^*$, and a minimum energy deposition for all calorimeters involved.

\begin{figure}
\includegraphics[trim={0.6cm 0.5cm 0.2cm 1.25cm},clip,width=\columnwidth]{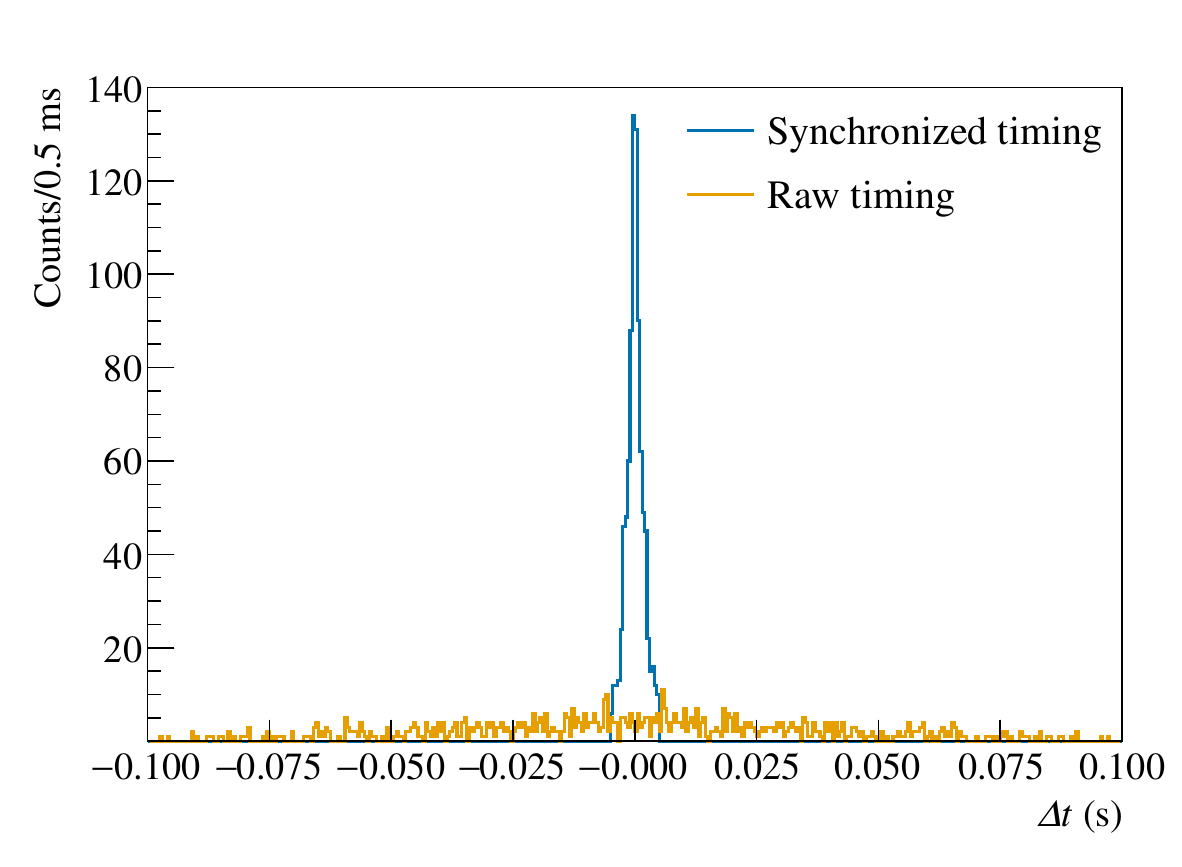}
\caption{Distribution of $\Delta t$ for double coincidence (M\,=\,2) 
events where the energies released in the two calorimeters sum to $\sim$2615\,keV. The distribution is obtained from calibration data of a single dataset and it is shown before (orange) and after (blue) the time response synchronization.}
\label{fig:jitter}      
\end{figure}

To optimize $\Delta t^*$, we first identify preliminary M\,=\,2 events that occur within 500\,ms of each other, where each event in the multiplet deposits an energy above 30\,keV, and with both energies summing to $\sim$2615\,keV. This ensures most events in the distribution correspond to true physical coincidences. As demonstrated in Fig.~\ref{fig:jitter} (in orange), the distribution of $\Delta t$ for these M2 events has a broad distribution, which also features fine structures. These structures are attributed to the characteristic rise time of individual calorimeters, which varies between 50\,--\,200\,ms, depending on the operational temperature of the detectors. For example, a simultaneous energy deposition in two calorimeters gives rise to a pair of events that have a characteristic time difference that is attributed to their different rise times. We refer to this characteristic time difference as the \emph{delay} between the two calorimeters.

To obtain a more precise $\Delta t$ among coincident events, we synchronize all calorimeters by correcting for the \emph{delay}; this allows us to use a narrower $\Delta t^*$. We use two different methods for calculating the \emph{delay}. The primary method uses coincident events, while the secondary method is used for designated reference calorimeters, or when the primary method is unavailable due to low statistics. Both methods are applied on a tower basis, with each tower having a designated reference calorimeter. All towers are then synchronized with respect to an overall reference calorimeter.

Using calibration data for high statistics, we identify M\,=\,2 events where the total energy deposited is $2615\,\pm\,25$\,keV. For adjacent calorimeter pairs, the first method calculates their \emph{relative delay} by averaging the $\Delta t$ of their multiplets. Instead of using physical coincidences, the second method uses pulser events to provide high-statistics sets of synchronized pulses. For each pulser event, an associated rise time is calculated as the time difference between the trigger position and the maximum of the waveform, as computed by the Optimum Filter (see Sec.~\ref{of}). For each calorimeter, the mean rise time of its pulser events is then used as a proxy for its characteristic rise time. Thus, the {\emph{relative delay}} between calorimeter pairs is calculated as the difference between their rise time proxies. If necessary, this method can be used for non-adjacent calorimeter pairs. 

After calculating all possible relative delays in the tower, we calculate the {\emph{absolute delay}}. Within a tower, the {\emph{absolute delay}} of a calorimeter is calculated by summing the {\emph{relative delays}} along the path of adjacent calorimeters, to the reference calorimeter, that minimizes the uncertainty on the {\emph{absolute delay}}. We finally synchronize each CUORE calorimeter with respect to the designated overall reference calorimeter by summing its absolute delay, rise time proxy of its reference calorimeter, and, if in another tower compared to the overall reference calorimeter, the rise time proxy of the overall reference calorimeter. The effect of the time response synchronization on the $\Delta t$ distribution of physical coincidences is demonstrated in Fig.~\ref{fig:jitter}; the time synchronized distribution (in blue) is the basis for using $\Delta t^*$\,=\,10\,ms for the \onbb decay analysis.

\subsection{Coincidence data}\label{CoincData}

For each event in a given multiplet, we compute and store the associated coincidence data: 
\begin{itemize}
    \item multiplicity
    \item DAQ channel number associated with the calorimeter 
    \item synchronized time stamp
    \item energy
    \item total energy from summing the energy of all events in the multiplet 
    \item index indicating the time ordering 
\end{itemize}

Coincidence events can be reconstructed in different ways, and depending on the physics process we are investigating, a fine-tuning of the coincidence parameters might be necessary. In addition to the coincidence window, $\Delta t^*$, parameters and options that can be applied to the coincidence algorithm that builds the multiplets include: 
\begin{itemize}
    \item \emph{Energy threshold}: the minimum energy an event must have to be considered for coincidence tagging. This can be tuned for each calorimeter, for example setting it to the OT threshold computed as described in Sec.~\ref{thresholds};
    \item \emph{Running window}: if this option is selected, the coincidence window will open at the first event in the multiplet and close when there are no more events within time $\Delta t^*$ of the last event. In this case the effective coincidence window can become arbitrarily long. 
    \item \emph{Distance cut}: the distance between triggered calorimeters represents an effective way to reduce inefficiencies due to accidental coincidences; it is unlikely that two simultaneous events that happen far apart are causally related. Two events, where one deposits energy in calorimeter $a$ and the other deposits energy in calorimeter $b$ will be considered coincident if 
    $$
        (\Delta t_{\mathrm{a,b}} \le \Delta t^*)\;\&\;(R_{\mathrm{a,b}} \le R_{\mathrm{max}})
    $$
    where R$_{a,b}$ is the distance between the centers of the corresponding crystals and R$_{max}$ is the value of the distance cut. The distance between any pair of CUORE crystal centers ranges between 60\,--\,1020\,mm. 
    \item \emph{Running radius}:  as in the case of the running window, if this option is selected, the coincidence window will open at the first event in the multiplet and close when there are no more events within distance $R_{\mathrm{max}}$ from any event in the multiplet. 
\end{itemize}
The value of the energy threshold for coincidence tagging is usually set to 40\,keV for physics analyses such as the \Te \onbb decay search or the CUORE background model (BM). For both these studies, we are mostly interested in the energy region above 100\,keV and we aim to reduce the rate of accidental coincidences in high rate calorimeters by setting a threshold. Consequently, we use a coincidence window of 10\,ms ($\pm$\,5\,ms) for the \onbb decay analysis, while a wider coincidence window of 60\,ms ($\pm\,30$\,ms) is used for the BM analysis and in general whenever there is the need to accurately calculate coincidences over a large energy range, from the gamma region (below 3\,MeV) to the alpha region. The time synchronization described in Sec.~\ref{TimeSync} allows us to implement a 10-ms coincidence window; it is optimized using M\,=\,2 coincidence events with total energy deposited by the multiplet $\sim\,$2615\,keV, which is close to Q$_{\beta\beta}$. A wider coincidence window is chosen for the BM analysis since we observe that the shape of the filtered pulse has a residual dependence on energy at the edges of the energy range of interest---i.e., for nuclear recoils and $\alpha$ particles---so by enlarging the coincidence window, we can safely neglect this dependence over a broad energy range. At the same time, the running window option is not selected for the BM analysis and only used for the \onbb decay analysis, while the running radius option is always applied with R$_{\mathrm{max}}\,=\,150$\,mm, which is related to the distance between two crystals facing each other across adjacent towers. Finally, the low energy analyses, which search for dark matter and axions, and focus on the energy region below 100\,keV, exploit the feature of calorimeter-dependent thresholds for the identification of coincident events.

\section{Pulse shape analysis}\label{sec:psd} 

To further reduce the background, we perform pulse shape discrimination (PSD) to identify and remove even\-ts associated with pulse shapes that are atypical compared to particle-induced events and thus more likely induced by unphysical events. 

\begin{figure}
  \includegraphics[trim={0cm 0cm 1.9cm 1cm},clip,width=\columnwidth]{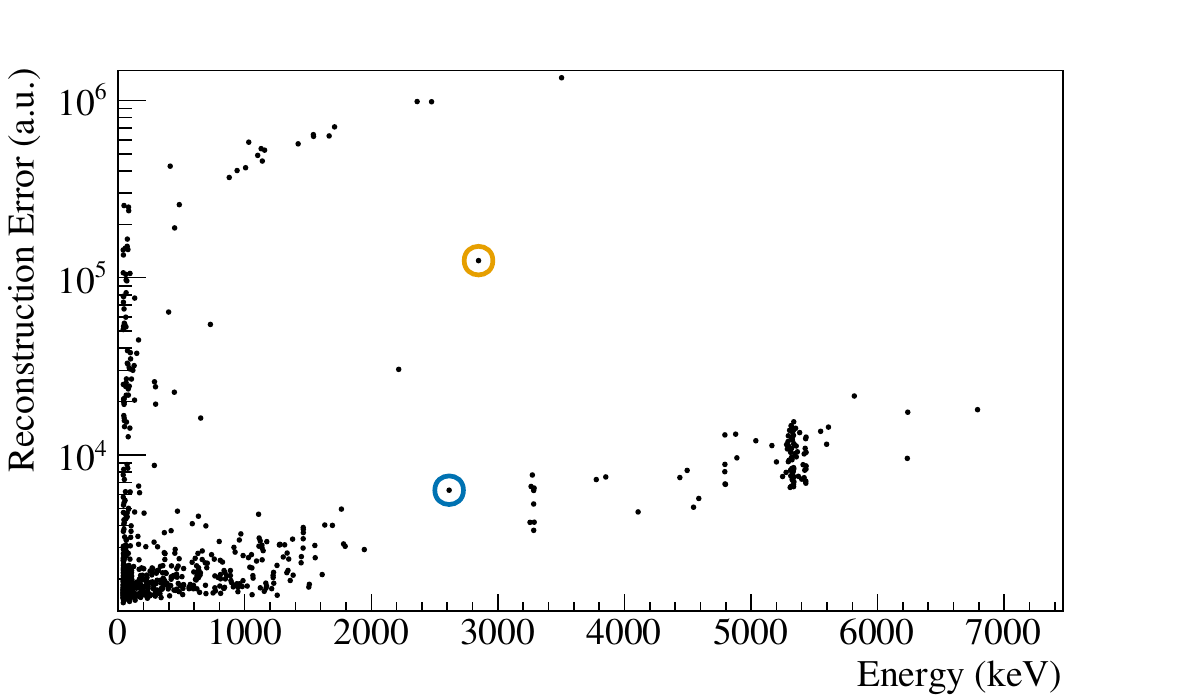}
\caption{Reconstruction error of M\,=\,1 events that passed basic quality cuts for a single Ch-DS. The blue circle indicates the data point associated with the accepted pulse shown in Fig.~\ref{fig:PCAgoodbadPulse} (top). The orange circle indicates the data point associated with the rejected pulse shown in Fig.~\ref{fig:PCAgoodbadPulse} (bottom).}
\label{fig:pcaRE}       
\end{figure}

\begin{figure}
  \includegraphics[trim={1.4cm 2.3cm 2cm 1cm},clip,width=\columnwidth]{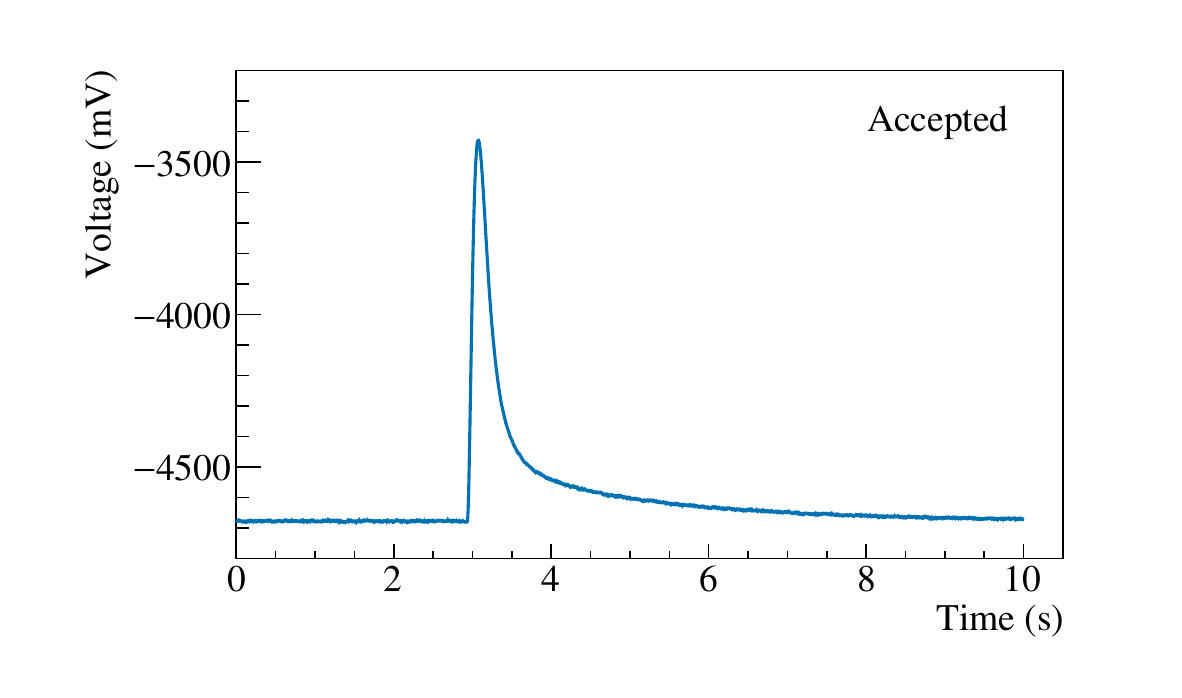}
  \includegraphics[trim={1.4cm 1.01cm 2cm 1cm},clip,width=\columnwidth]{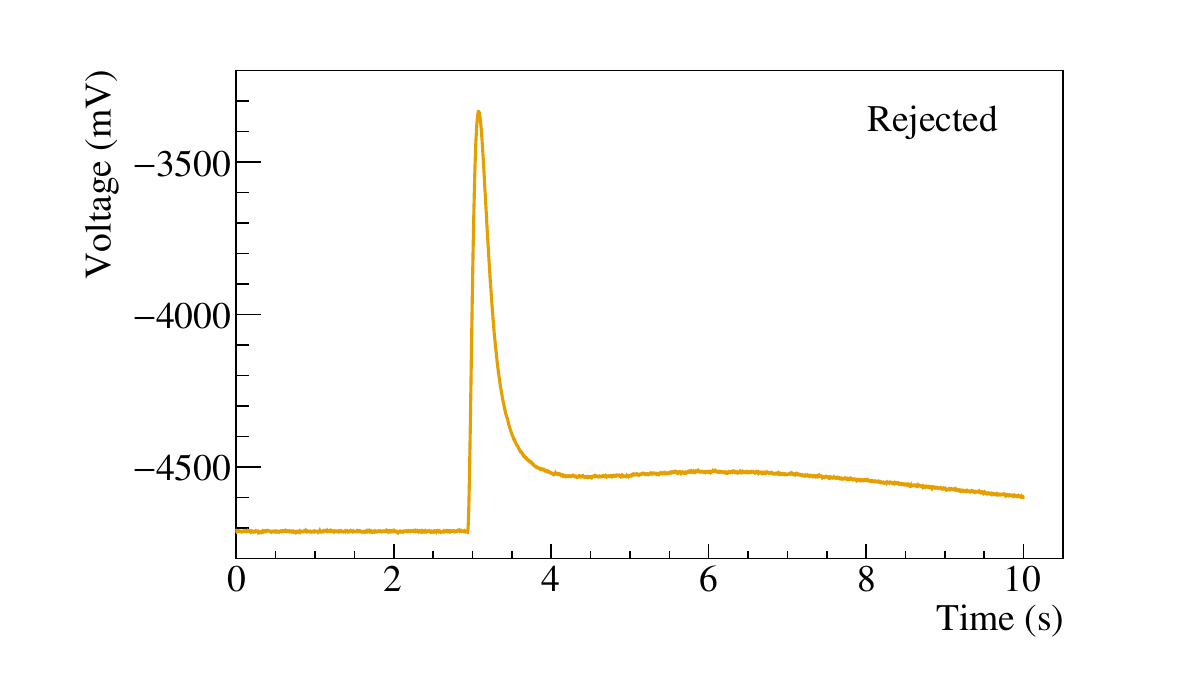}
\caption{Top: Example of an accepted pulse with 2615\,keV energy and RE\,=\,6328\,a.u. (NRE\,=\,3.6 a.u.). Bottom: Example of a rejected pulse with 2847\,keV energy and RE\,=\,124817 a.u. (NRE\,=\,228.4\,a.u.).}
\label{fig:PCAgoodbadPulse}       
\end{figure}

The Reconstruction Error method is used to discriminate between physical and non-physical (pileup, noise, etc.) events. 
This method compares each event ($\boldsymbol{ v}$) with an average pulse ($\boldsymbol{ v_{AP}}$) that is representative of a particle-induced pulse in an energy range between $\sim$100\,keV and $\sim$2615\,keV~\cite{Huang_PhD-thesis:2021}. We calculate the Reconstruction Error (RE) as
\begin{equation} \label{eq:RE}
     \mathrm{RE} = \sqrt{\sum_{d=1}^D\Big[v_d - \big(\boldsymbol{ v} \cdot \boldsymbol{v}_{AP} \big) v_{AP,d} \Big]^2},
\end{equation}
where $d$ is the sample index in the waveform. In this form, the RE quantifies the difference in pulse shape between a signal-triggered event and its corresponding average pulse. 
Figure~\ref{fig:pcaRE} shows the reconstruction errors of M\,=\,1 events that passed basic quality cuts. The lower diagonal band represents signal-like events, which distinctly include physical alpha events from $^{190}$Pt and $^{210}$Po above 3\,MeV. A pulse corresponding to an event inside and outside the signal-like band is shown at the top and bottom of Fig.~\ref{fig:PCAgoodbadPulse}, respectively.  

We normalize the RE distribution by fitting the signal-like band with a second-order polynomial $f(E)$, and computing the median absolute deviation (MAD) with respect to the best-fit curve. The normalized reconstruction error (NRE) is then defined as
\begin{equation}
\label{eq:2}
   {\textrm{NRE}} = \frac{{\textrm{RE}}-f(E)}{\textrm{MAD(E)}}\textcolor{blue}{.}
\end{equation}

\begin{figure}
  \includegraphics[trim={0cm 0cm 1.9cm 1.2cm},clip,width=\columnwidth]{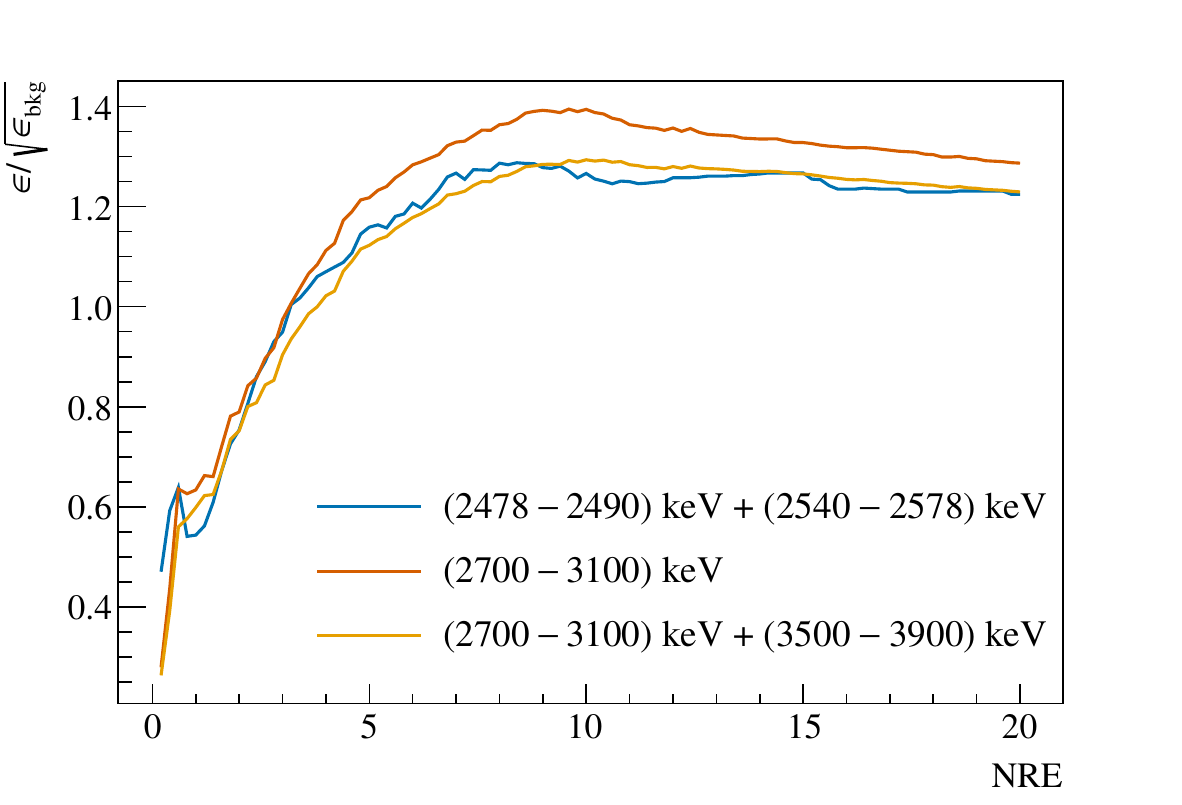}
\caption{$\mathrm{FOM_{PSD}}$ optimization curves from combining two datasets based on the selection efficiency in the 2615\,keV peak, $\epsilon$, and the selection efficiency in the background region, $\epsilon_{\mathrm{bkg}}$. The background region(s) used to calculate $\epsilon_{\mathrm{bkg}}$ are indicated in the legend. Only the orange curve is used to decide the NRE threshold, however we present the blue and yellow curves to show that using alternative background regions for $\epsilon_{\mathrm{bkg}}$ would yield the same result.}
\label{fig:PSDFOM}  
\end{figure}

In order to select good quality physics pulses, we apply a cut on the NRE value. We identify the NRE threshold empirically by optimizing the following figure of merit:  
\begin{equation}
\label{eq:FOM}
    \mathrm{{FOM}_{PSD}} = \frac{\epsilon}{\sqrt{\epsilon_{\mathrm{bkg}}}}
\end{equation}
where $\epsilon$ ($\epsilon_{\mathrm{bkg}}$) is the efficiency of accepting signal (background) events within a predefined energy region, taken as 2615\,$\pm$\,10\,keV (2700\,--\,3100\,keV). 
In order to avoid biasing our decision on the upper NRE threshold, for each dataset, the $\mathrm{FOM_{PSD}}$ is calculated using half of the data, selected at random. Based on this procedure, we set an NRE threshold of 10 for all calorimeters across all datasets. Figure~\ref{fig:PSDFOM} demonstrates the typical trend of the $\mathrm{FOM_{PSD}}$ as a function of the normalized reconstruction error cut value. 

\section{Blinding}
\begin{figure}
  \includegraphics[trim={0.25cm 0cm 1.4cm 1.25cm},clip,width=\columnwidth]{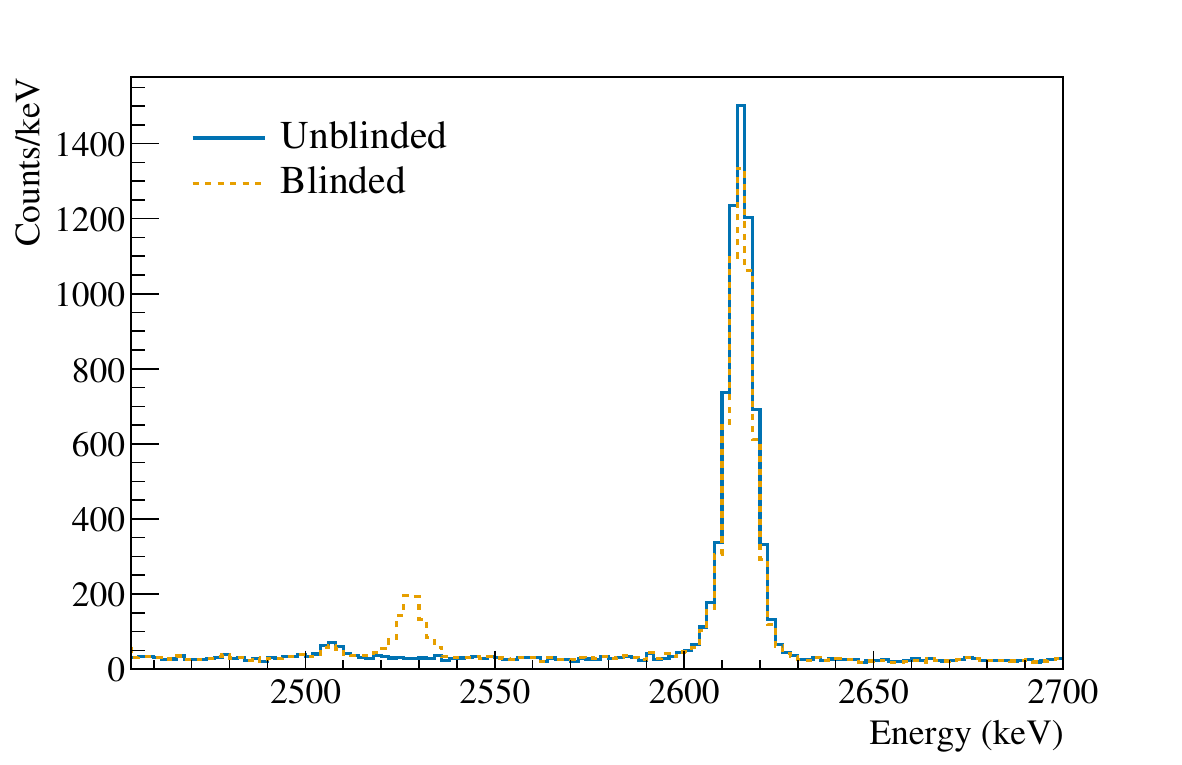}
\caption{Comparison of blinded (salted peak at $Q_{\beta\beta}$) and unblinded spectra for the total 2\,\tony data release after all \onbb decay analysis cuts.} 
\label{fig:peakSalting}      
\end{figure}

A data salting procedure, which consists of inserting an artificial peak at $Q_{\beta\beta}$, is applied to the \onbb decay region of interest (ROI). This peak is constructed by exchanging a random fraction (10\,--\,12.5$\%$) of events between the 2615\,keV peak and the posited $Q_{\beta\beta}$. The chosen fraction range ensures the fake peak in the data would camouflage a potential true signal peak, whose amplitude cannot be greater than the limit from the previous Te-based experiments. Specifically, events within $\pm$\,50\,keV of either energy are randomly shifted by the difference between the two peak energies, $|E(^{208}\mathrm{Tl})-Q_{\beta\beta}|$. Figure~\ref{fig:peakSalting} shows a comparison of the blinded and unblinded \onbb decay spectra in the relevant energy range. A similar data salting procedure is applied in other peak-searches, such as \onbb decay of $^{128}$Te~\cite{Adams:2022hji} or $^{120}$Te~\cite{CUORE:2022dwv}, to blind their respective ROIs.

\section{Detector energy response for \onbb decay} 

In order to search for the potential \onbb decay signal in the data, we model the ROI as a flat background with peaks at the posited $Q_{\beta\beta}$ value and $^{60}$Co sum peak energy, where the latter originates from the simultaneous deposition of two $^{60}$Co gamma rays (at 1173\,keV and 1332\,keV) in the same crystal. This section describes our peak shape characterization and the evaluation of the detector energy response in the \onbb decay ROI. Specifically, we determine the energy-dependent resolution and energy bias, where the latter is defined as the difference between the measured peak position and nominal characteristic gamma-ray energy, and extrapolate them to $Q_{\beta\beta}$ for each dataset and active calorimeter.

\subsection{Lineshape} 

The general peak shape, interchangeably described as lineshape, is modeled after the $^{208}$Tl 2615\,keV photopeak from calibration data for each Ch-DS. We use calibration data instead of physics data to ensure high statistics and reliable detector response fit parameters for most Ch-DS pairs. 

Our phenomenological lineshape model of a photopeak consists of a sum of up to 3 Gaussians--a main Gaussian and two lateral Gaussian sub-peaks positioned at the left (L) and the right (R) of the main one---all sharing the same width~\cite{cuore2025:science}:
\begin{multline}
    f_{cal}(E;\mu_{c},\sigma_{c},A_L,A_R,a_L,a_R) \doteq \\
    \frac{\mathcal{G}(E;\mu_{c},\sigma_{c}) + A_L \mathcal{G}(E;a_L \mu_{c},\sigma_{c}) + A_R \mathcal{G}(E;a_R \mu_{c},\sigma_{c})}{ 1 + A_L + A_R}
    \label{eq:LineshapeFormula}
\end{multline}
where $\mathcal{G}(E;\mu_c,\sigma_c)$ indicates the main Gaussian function centered at $\mu_c$ with standard deviation $\sigma_c$, while $A_{L/R}$ and $a_{L/R}$ scale the amplitudes and positions of the Gaussian sub-peaks, respectively. We perform the fit with a series of iterations, setting $A_{L/R}=0$ in cases where $A_{L/R} < 10^{-6}$ and broadening the fit parameter ranges in cases of railing.  

\begin{figure}
  \includegraphics[trim={0.59cm 0cm 0.85cm 0.6cm},clip,width=\columnwidth]{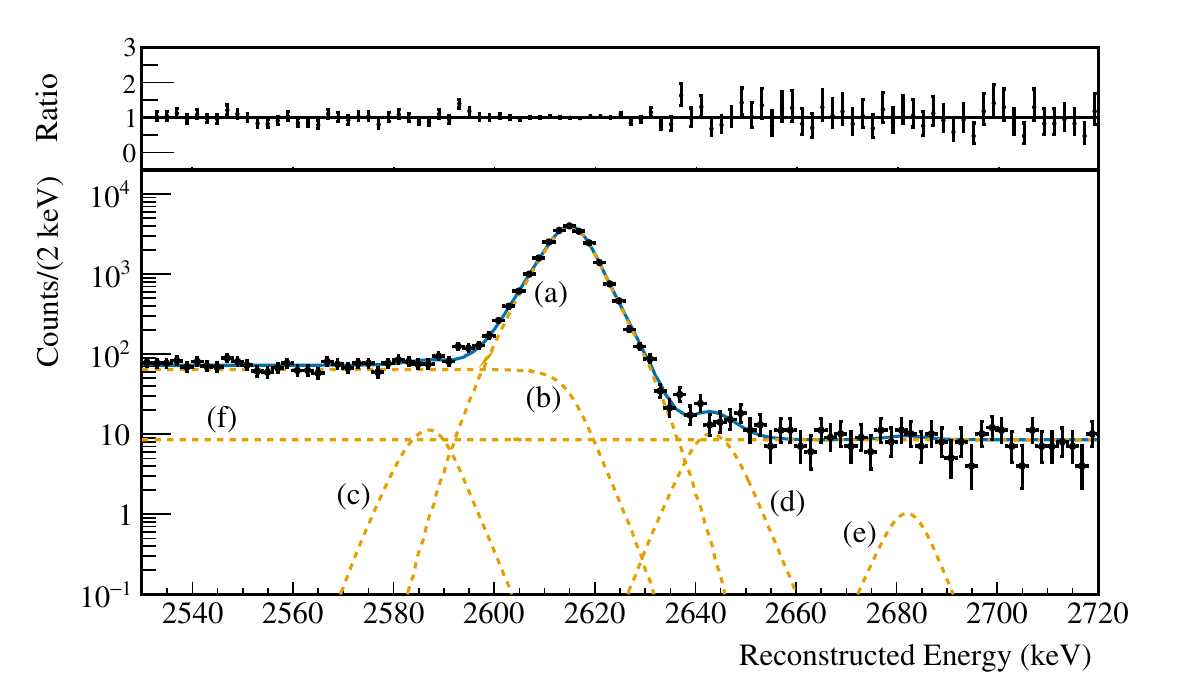}
\caption{Top: Ratio between calibration data and peak shape model. Bottom: Summed fit result of the $^{208}$Tl peak in calibration data for a single tower-dataset. We label the different fit components: (a) $^{208}$Tl photopeak, (b) multi-Compton continuum, (c) 2615\,keV $\gamma$ with $^{130}$Te X-ray escape, (d) 2615\,keV $\gamma$ and $^{130}$Te X-ray coincidence (e) 2615~keV and $^{208}$Tl 583\,keV $\gamma$ coincidence with e$^+$/e$^-$ 511\,keV escape, and (f) uniform background. This figure was adapted from~\cite{cuore2025:science}.}
\label{fig:LScalib}   
\end{figure}

The lineshape model adapts to the $^{208}$Tl 2615\,keV calibration photopeak, however, to properly account for the features of the calibration spectrum between 2530\,--\,2720\,keV, we also include the following components in the fit model~\cite{Alduino:2016zrl,Huang_PhD-thesis:2021}:
\begin{itemize}
    \item a continuum that extends into the left side of the peak and is due to 2615\,keV $\gamma$s undergoing multi-Compton scattering 
    \item a peak at 2585\,keV resulting from the photo-absor\-ption of a 2615\,keV $\gamma$ followed by the escape of a $^{130}$Te X-ray whose energy ranges between 27\,--\,31\,keV
    \item the coincidence of a 2615\,keV $\gamma$ and a $^{130}$Te X-ray
    \item a peak at 2687\,keV due to the summed energy of $^{208}$Tl 583\,keV + 2615\,keV $\gamma$s followed by a e$^+$/e$^-$ (annihilation) 511\,keV escape 
    \item a flat background 
\end{itemize}
While the lineshape parameters of the photopeak are defined at a Ch-DS level, due to limited statistics, we define the parameters to characterize the additional spectral features at tower-dataset level and perform the fit simultaneously over all calorimeters in a given tower-dataset. Figure~\ref{fig:LScalib} shows a composite fit for a single tower-dataset. With respect to this procedure, on average we exclude $\sim$60 calorimeters per dataset from this and further analysis steps due to poor statistics. 

\subsection{Energy scaling} 

We use physics data to determine the energy dependence of the lineshape parameters, $\sigma$ and $\mu$, for each Ch-DS. Due to the limited statistics, we fit the spectrum of each dataset, summed over all channels. We assume that the resolution term can be described as the quadrature sum of an energy-dependent term, $\sigma_E$, and an energy-independent term, $\sigma_\circ$, that depends on the intrinsic features of the detectors: 
\begin{equation}
    \sigma(E) = \sqrt{(\sigma_E)^2 + (\sigma_{\circ})^2},
\end{equation}
and that the energy-dependent term scales across phy\-sics data in a way that is proportional to calibration data. Thus, in physics data we model the energy resolution, $\sigma_p$, of a given peak at nominal energy $E$ as  
\begin{equation}
\label{eq:res}
    \sigma_{p}(E) = \sqrt{\sigma^2_{p,\circ} + R_\sigma(E) \cdot (\sigma^2_c - \sigma^2_{c,\circ})}
\end{equation}
where $\sigma_{p,\circ}$ ($\sigma_{c,\circ}$) is the baseline resolution extracted using noise events on a Ch-DS basis in physics (calibration) data and $R_\sigma(E)$ is a dataset-level fit parameter that completely isolates the energy dependence. The peak position, $\mu_p$, is parameterized as 
\begin{equation}
\label{eq:bias}
    \mu_{p}(E) = R_\mu(E) \cdot E
\end{equation}
where $R_\mu$ is a dataset-level fit parameter. For each peak, the relative amplitudes and relative positions of the Gaussian subpeak(s) of each Ch-DS are fixed to the values extracted from the calibration fit.

\begin{table}[h!]
    \centering
    \begin{tabular}{cc}
\hline
\textbf{Source}    & \textbf{Q-value (keV)} \\ 
\hline
$^{208}$Tl	                    & 2614.5         \\
$^{40}$K	                    & 1460.8         \\
$^{228}$Ac	                    & 911.2         \\
$^{214}$Bi	                    & 1120.3         \\
$^{60}$Co	                    & 1173.2         \\
$^{60}$Co	                    & 1332.5         \\
$^{214}$Bi	                    & 1238.1         \\
$^{214}$Bi	                    & 1764.5         \\
$^{214}$Bi	                    & 2204.0         \\
$^{208}$Tl	                    & 583.2         \\
$^{214}$Bi	                    & 1729.6         \\
$^{228}$Ac	                    & 338.3         \\
$^{228}$Ac	                    & 969.0         \\
$^{212}$Pb	                    & 238.6         \\
$^{228}$Ac	                    & 1588.1         \\
$^{212}$Bi	                    & 727.3         \\
$^{54}$Mn	                    & 834.8         \\
$^{214}$Bi	                    & 2447.9         \\
$^{125m}$Te	                    & 144.8         \\
$^{214}$Pb	                    & 351.9         \\
\hline
\end{tabular}
    \caption{Subset of characteristic gamma lines in physics data identified by our background model. }
    \label{tab:PeakID}
\end{table}

\begin{figure}
  \includegraphics[width=\columnwidth]{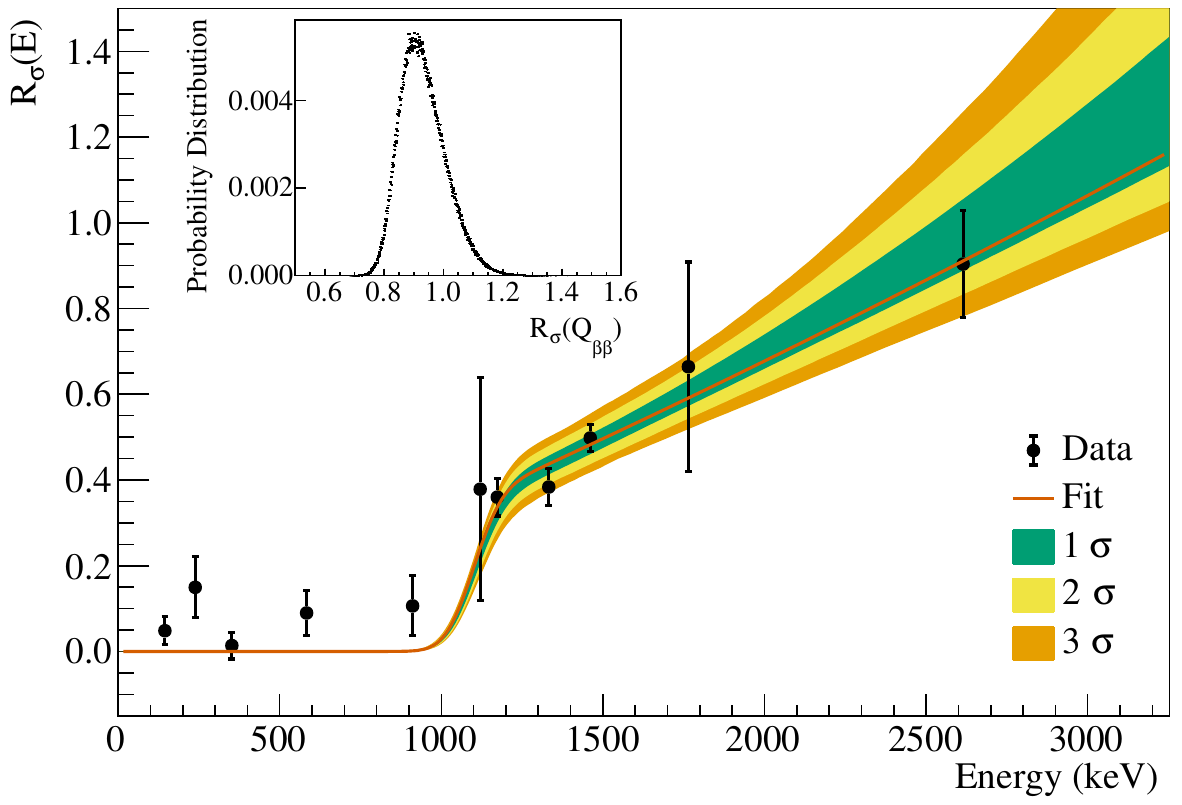}
  
\caption{Single dataset fit of the energy-dependent resolution scaling term R$_\sigma$ from physics data. Inset: Marginalized posterior for R$_\sigma$(Q$_{\beta\beta}$) for the corresponding dataset.}
\label{fig:ResoScale}  
\end{figure}

Table \ref{tab:PeakID} lists the peaks we consider in order to determine $R_\sigma(E)$ and $R_\mu(E)$. Among them, we select peak fits with adequate statistics and reasonable convergence and use them to evaluate the energy dependence of the detector response. From the selected subset of peaks, for each dataset, we fit $R_\sigma(E)$ to a phenomenological function that trends as a quadratic polynomial near $Q_{\beta\beta}$, using the BAT~\cite{BATSoftware} package. This function is forced to zero at low energies to avoid unphysical, negative values; at higher energies, $R_{\sigma}(E)$ is quadratic. To transition between the two energy regions, a sigmoid-like rolling function centered at $\sim$1000\,keV is used; the full fit function is shown in Fig.~\ref{fig:ResoScale}. The marginalized posterior distribution for $R_\sigma$ from each dataset is then used as a prior in the \onbb decay ROI fit. This allows us to maintain a reasonable number of parameters in our fit. The inset of Fig.~\ref{fig:ResoScale} shows the marginalized posterior for $R_\sigma$ at $Q_{\beta\beta}$ for a single dataset. 

\begin{figure}
  \includegraphics[width=\columnwidth]{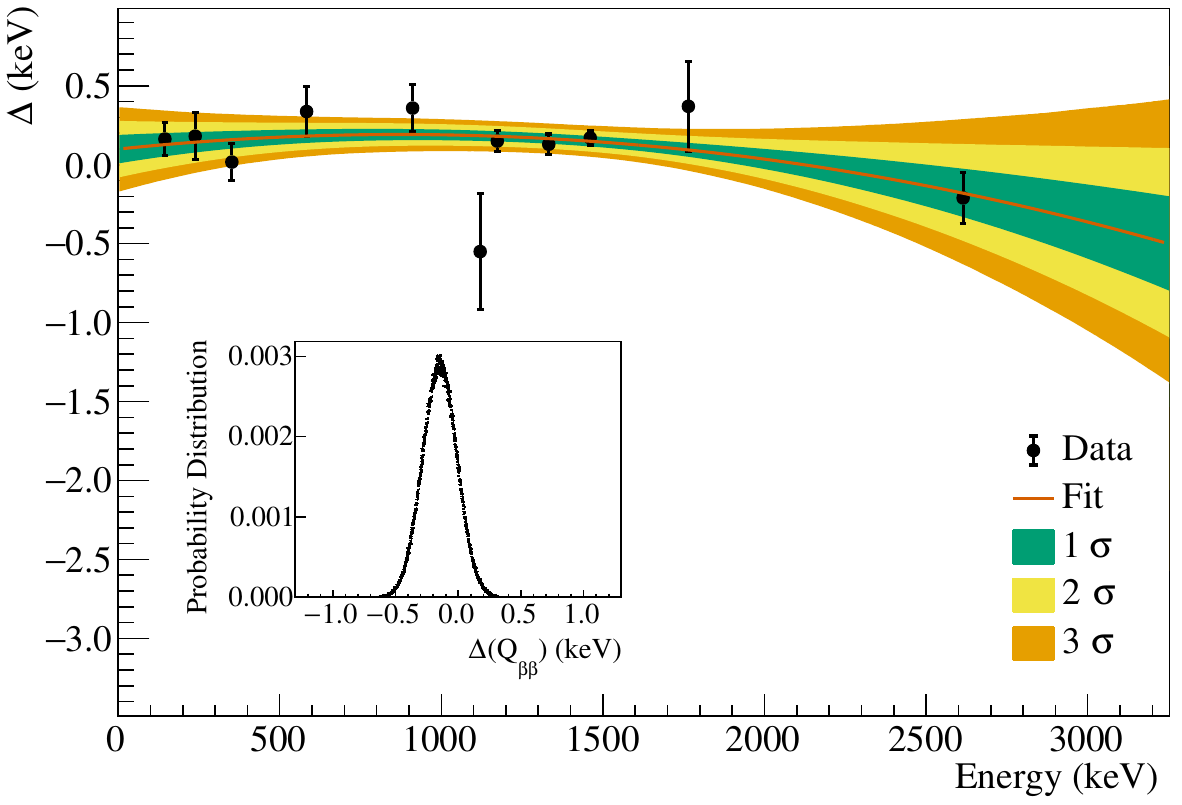}
\caption{Single dataset fit of the energy bias $\mathrm\Delta$. Inset: Marginalized posterior for the energy bias at Q$_{\beta\beta}$ for the corresponding dataset.}
\label{fig:BiasScale}  
\end{figure}

We fit the bias on the peak position, defined as 
\begin{equation}
\label{eq:actbias}
    \mathrm\Delta(E) = \mu_p(E) - E
\end{equation}
to a quadratic polynomial in BAT. 
Figure~\ref{fig:BiasScale} shows the energy dependence of the energy bias for a single dataset. The inset of Fig.~\ref{fig:BiasScale} shows the corresponding marginalized posterior for the bias at $Q_{\beta\beta}$ for the same dataset. When fitting the ROI, we model the expected position of the \onbb decay peak in our data for a given Ch-DS as 
\begin{equation}
\label{eq:Qfit}
    Q_{fit} = \frac{\mu_c}{2615\,\mathrm{keV}}Q_{\beta\beta} + \mathrm\Delta(Q_{\beta\beta}).
\end{equation}
In the first term, the ratio is between the measured peak position ($\mu_c$) from Eq.~\ref{eq:LineshapeFormula} and its nominal value, while $Q_{\beta\beta}$ is a global floating parameter which is sampled from a Gaussian prior probability distribution for $Q_{\beta\beta}$; the ratio is used to compensate for any possible misconstruction of the peak position from the Ch-DS dependent calibration fit at 2615\,keV. The second term is a dataset- and channel-dependent floating parameter that accounts for the energy bias determined from the physics data. Namely, $\mathrm\Delta(Q_{\beta\beta})$ is a random sample from the multivariate prior probability distribution for the energy bias at $Q_{\beta\beta}$. 

\section{Exposure and efficiency}
The exposure is a crucial parameter that is used to quantify the experimental sensitivity---the potential to detect rare interactions. This quantity is typically defined as the product of the active mass of the detector and the time over which data used for the search is collected. From the collected exposure, we identify and remove data unsuitable for analysis and distinguish the remaining data as the analyzed exposure. In our analysis, there are two main categories of quality cuts: time-based cuts and event-based cuts. The former impact the final value of the exposure (see Sec.~\ref{sec:expo}), while the latter are used in our selection efficiency evaluations (see Sec.~\ref{sec:bc}\,--\,Sec.~\ref{sec:psdeff}).

\subsection{Time-based cuts} \label{sec:IntervalCuts}
At the beginning of the data production, it is necessary to identify and reject \emph{bad intervals}, which flag time periods when the entire detector or a subset of calorimeters experience disturbances in the data taking conditions. These calorimeter-intervals are excluded from the analysis since they may not be representative of the behavior of the calorimeter(s) over the dataset. This is performed manually on the CUORE Online Run Control (CORC) system, a diagnostic system developed to monitor all calorimeters by providing an overview of their performance over time. 

\emph{Bad for analysis} is another type of flag that is used to indicate that data processing cannot proceed. For example, the initial and final few tens of seconds of each run are flagged since denoising cannot reliably be applied. Calorimeters are cut from the entire dataset when sequences fail to compute the parameters necessary for the analysis---e.g., parameters from stabilization, calibration, lineshape.

\subsection{Pulser crosstalk exposure correction} 
In CUORE, heater events are driven by a pulser board injecting signals on a set of detector columns at the same time~\cite{PulserCUORE}. Due to electrical interference, this can induce \emph{crosstalk} pulses in other detector channels outside of the expected set. Channels affected by this 
issue are effectively blind to good physics data when a crosstalk pulse is present. We exploit the known periodicity of the pulser cycle to quantify the extent of crosstalk and calculate a corresponding exposure correction.

This statistical analysis relies on three samples of events: pulser events, \emph{all} signal events, and \emph{good} signal events---selected after basic quality cuts. For each calorimeter, using the synchronized pulser trigger times, we consider the distribution of events within a time window that spans the pulser trigger, and another time window that is generally free of crosstalk. The former is the crosstalk window, taken as $\pm$1 second of the pulser firing time, while the latter is the control window, taken as a 1 second interval preceding the crosstalk window. For each calorimeter, we use the \emph{all} signal events in the control window to estimate the expected signal events, $n_{control}$. The \emph{good} signal events in the crosstalk window are used to estimate the number of events tagged as good pulses ($n_{tag}$). This sample includes both crosstalk events and particle pulses, so the probability of a crosstalk event occurring from a given pulser column firing, {$P_{x,col}$ can be expressed as
\begin{equation}
    P_{x,col} = \frac{n_{tag} - n_{control}}{N_{PulserCycles}}
\end{equation}
where $N_{PulserCycles}$ is the number of pulser cycles considered. The impact of the pulser crosstalk on the total exposure of a given channel, $\zeta_{ch,tot}$, is the ratio of the corrected and total exposures:  
\begin{equation}
    f_p = \frac{\zeta_{ch,corr}}{\zeta_{ch,tot}} = \frac{1}{T_{P}} \sum_{col}  P_{x,col} \cdot \delta t_{col} 
\end{equation}
where $T_{P}$ is the period for firing on all detector columns, $\delta t_{col}$ is the time interval for firing subsequent sets of columns, and the summation is over all the CUORE tower columns. For any given dataset, only $\sim$20 channels are significantly affected by the pulser crosstalk in their exposure with $f_p$\,$>$\,10\%.

\subsection{Collected and analyzed exposure} \label{sec:expo}
The data quality selections, signal processing and analysis chain, in particular the bad intervals and bad analysis flags, reduce the collected exposure by $\sim$9\%, while the pulser crosstalk has an additional $\sim$1\% effect. For a total collected exposure of 2264.32\,\kgyn, the final analysis exposure is 2039.0\,\kgy of \teodn~\cite{cuore2025:science}. In Fig.~\ref{fig:expDS}, the collected and analyzed exposures for each dataset are reported. The average collected exposure across datasets is $\sim$80\,\kgyn; dataset number 5 has a much lower exposure due to cryogenic issues that caused an earlier stop of the data-taking. The exposure loss after the analysis cuts among datasets ranges between 6\,--\,15\%.

\begin{figure}
    \centering
\includegraphics[trim={0.45cm 0cm 1.7cm 1cm},clip,width=\columnwidth]{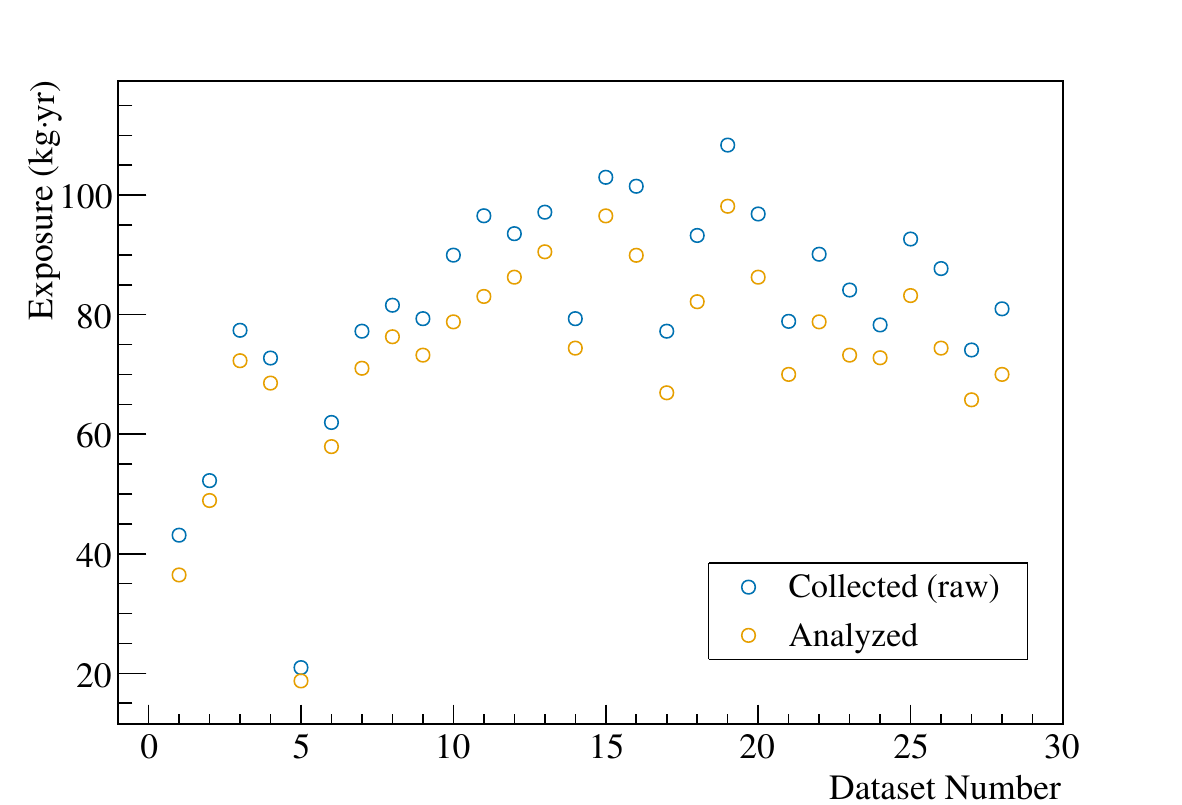}
    \caption{
    Raw exposure collected and exposure after analysis cuts (analyzed exposure) for each dataset included in the 2\,\tony data release.}
    \label{fig:expDS}
\end{figure}

To calculate the exposure, $\zeta$, in the $^{130}$Te isotope, we consider its isotopic abundance and the fraction of the mass of $^{130}$Te in the \teod crystals:  
\begin{equation}
    \zeta_{^{130}\mathrm{Te}} = \zeta_{\mathrm{TeO}_2} \times \frac{m.m.(\mathrm{TeO}_2)}{m.m.(^{130}\mathrm{Te})} \times \eta_{^{130}\mathrm{Te}}
\end{equation}
where $m.m.(\mathrm{TeO}_2)$ = 159.6\,g$\cdot$mol$^{-1}$ and $m.m.(^{130}\mathrm{Te})$ = 129.9\,g$\cdot$mol$^{-1}$ are the molar masses of \teod and $^{130}$Te compounds, respectively, and $\eta_{^{130}\mathrm{Te}}$ = 34.167\% is the natural abundance of $^{130}$Te. Therefore, the final analysis exposure in \teod corresponds to 567.0\,\kgy in $^{130}$Te.

\subsection{Base cuts efficiency} \label{sec:bc}  
The goal of the basic data quality cuts is to reject spurious events (baseline instabilities, noisy events, electric spikes, pileup events, etc.), while accepting pure single-site signal events. The efficiencies of these cuts are measured using heater pulses of known amplitude. The pulser-based base cuts efficiencies consist of three terms: detection, pileup rejection, and energy reconstruction.

\subsubsection{Detection efficiency}
The detection efficiency $\epsilon_{Det}$ measures the efficiency of the triggering algorithm. Every time a heater injects a given amount of power, a dedicated pulser flag is associated with the event; simultaneously, the triggering algorithm runs over the waveforms. The efficiency is evaluated as the ratio of the number of pulser events that are recognized by the trigger algorithm $N_{P}^{Trg}$ to the total number of injected pulser signals $N_{P}$:  
\begin{equation}
    \centering
    \epsilon_{Det} = \frac{N_{P}^{Trg}}{N_{P}}
\end{equation}

\subsubsection{Pileup rejection efficiency}
The pileup rejection efficiency $\epsilon_{PU}$ measures the efficiency of discriminating events that exhibit pileup, or two or more events occurring in the same 10-s event window. These can be multiple signal events or overlapping signal and heater events. The probability of not having a particle event with pile-up 
on the main pulse 
is related to two individual probabilities: the probability that, within the event window
\begin{itemize}
\item there is no second signal, and
\item there is no heater pulse. 
\end{itemize}
So the pileup rejection efficiency can be expressed as the product of two terms: 
\begin{equation}
    \centering
    \epsilon_{PU} = \frac{N_{P}^{Trg, noPileUp}}{N_{P}^{Trg}} \times \frac{(T_{P}-w)}{T_{P}}
\end{equation}
The first term indicates the number of triggered pulser events that are identified as single pulses. The second term is the probability that a pulser will not be injected into a signal event window; this depends on the size of the event window ($w$ = 10\,s) and the pulser firing period ($T_{P}$). Meanwhile, the probability of not having a pulser signal simultaneously with a particle event signal, when $w$ $\ll$ $T_{P}$, can be approximated as 
\begin{equation}
    \centering
    P(noPulser|IsSignal) \simeq 1 - \frac{w}{T_{P}}
\end{equation}

\subsubsection{Energy reconstruction efficiency}
The energy reconstruction efficiency $\epsilon_{EnRec}$ measures the efficiency of events being reconstructed with the correct energy, based on the energy reconstruction of pulser events with known energy. For a given pulser amplitude, events which are reconstructed with an energy within 3\,$\sigma$ of the mean are identified as well reconstructed. The ratio of these events to the total number of events in a window of [mean $\pm$\,10\,$\sigma$] is then taken as the efficiency of the energy reconstruction: 
\begin{equation}
     \centering
     \epsilon_{EnRec} = \frac{N_{P}^{3\sigma}}{N_{P}^{10\sigma}}
\end{equation}
The 10\,$\sigma$ window was chosen in order to avoid counting pulser events in which, due to the energy sequence failures, the energy is either reconstructed at zero or at the default value for the energy variable. 

\subsubsection{Combined base cuts efficiencies}
The three base cuts selection efficiencies described above are calculated for each channel-run (see Fig.~\ref{fig:EffRuns}). Then for each of the three base cuts, the DS-Ch efficiency is calculated as 
\begin{equation}
\epsilon_{\mathrm{BC,DS-Ch}}= \sum_{run}^{} n_{pass} \Big/ \sum_{run}^{}(n_{pass}\,+\,n_{fail}) 
\end{equation}
where $n_{pass}$ and $n_{fail}$ are the number of pulser events that pass and fail the base cut as previously defined, respectively. The dataset efficiency for each base cut is calculated by averaging over all channels with active heaters, weighting by exposure (see Fig.~\ref{fig:EffDss}). In general, the detection and energy reconstruction efficiencies for pulses in the \onbb decay ROI are nearly 100\%. The pileup rejection efficiency is slightly lower, $\sim$96\%, with mild variation along the datasets for a given channel, and among the datasets for the whole group of channels. This is related to variations of the noise with time. To evaluate a single base cut efficiency for the dataset, these three efficiencies are combined numerically. Namely, we construct a distribution for each one by using their mean and standard deviation, then take a random sample of each distribution and calculate their product. We build a probability distribution for the final base cut efficiency from these products. 
\begin{figure}
\includegraphics[trim={0.1cm 0cm 1.8cm 1cm},clip,width=\columnwidth]{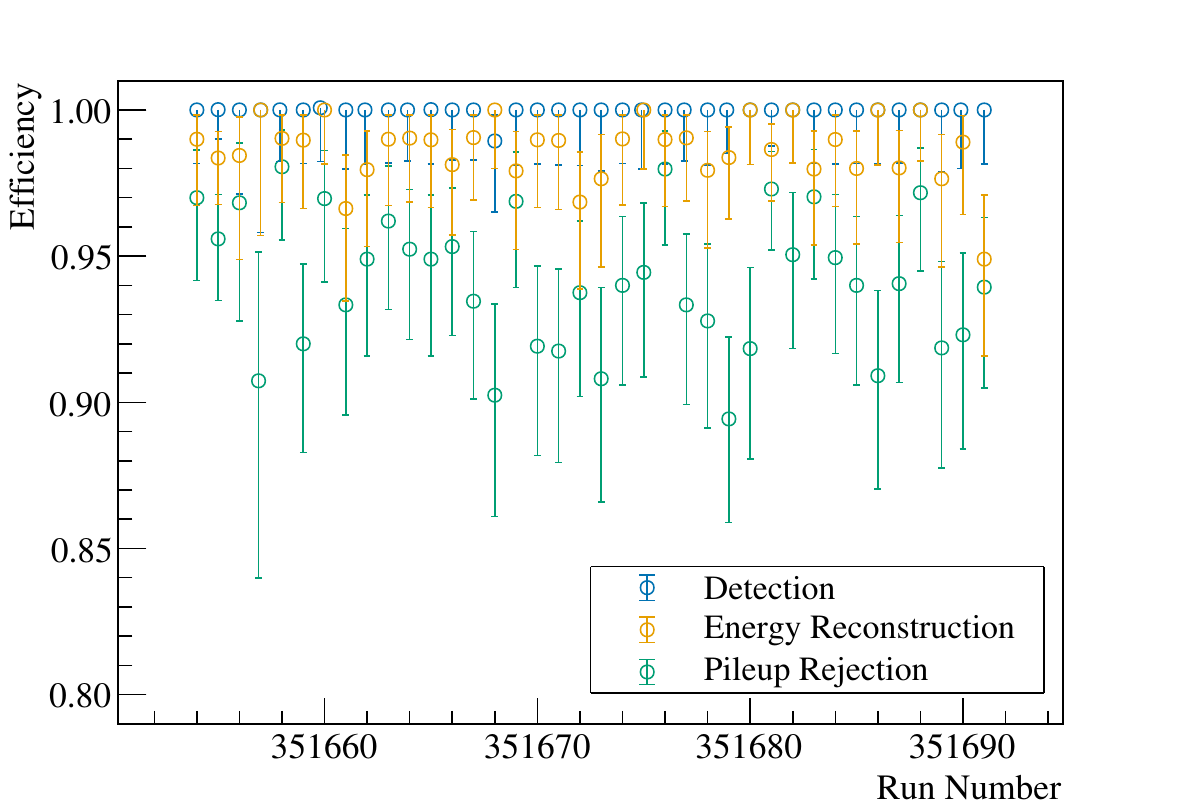}
\caption{Base cuts efficiencies for a single Ch-DS over sequential physics runs in a single dataset.} 
\label{fig:EffRuns}  
\end{figure}

\begin{figure}
\includegraphics[trim={0.1cm 0cm 1.8cm 1cm},clip,width=\columnwidth]{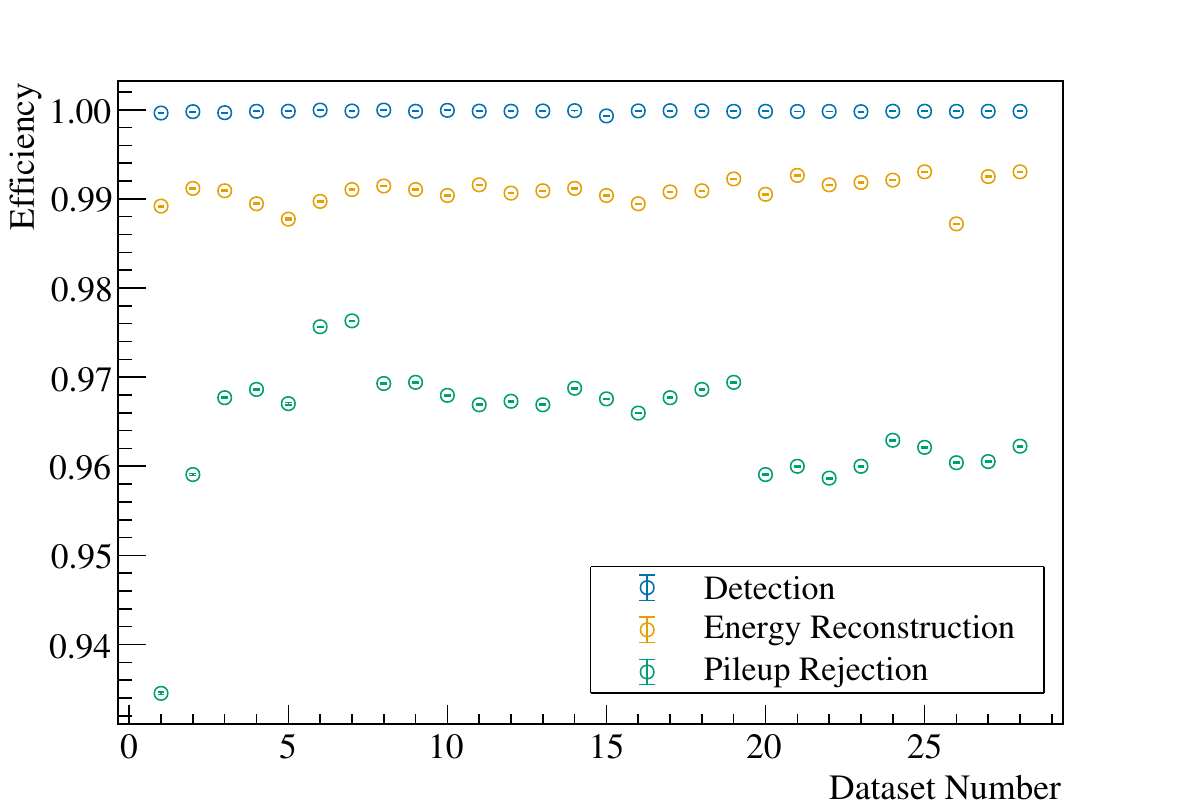}
\caption{Base cuts efficiencies averaged over the heater-active channels for the datasets included in the 2\,\tony data release. Note that the error bars are small and inside the open circle markers. The relatively lower pileup rejection efficiency of the first dataset is a consequence of the shorter pulser firing period that was initially used.} 
\label{fig:EffDss}      
\end{figure}

\subsection{Anti-coincidence efficiency} \label{sec:aceff}

\begin{figure}
  \includegraphics[trim={0.5cm 0.6cm 0.9cm 1.2cm},clip,width=\columnwidth]{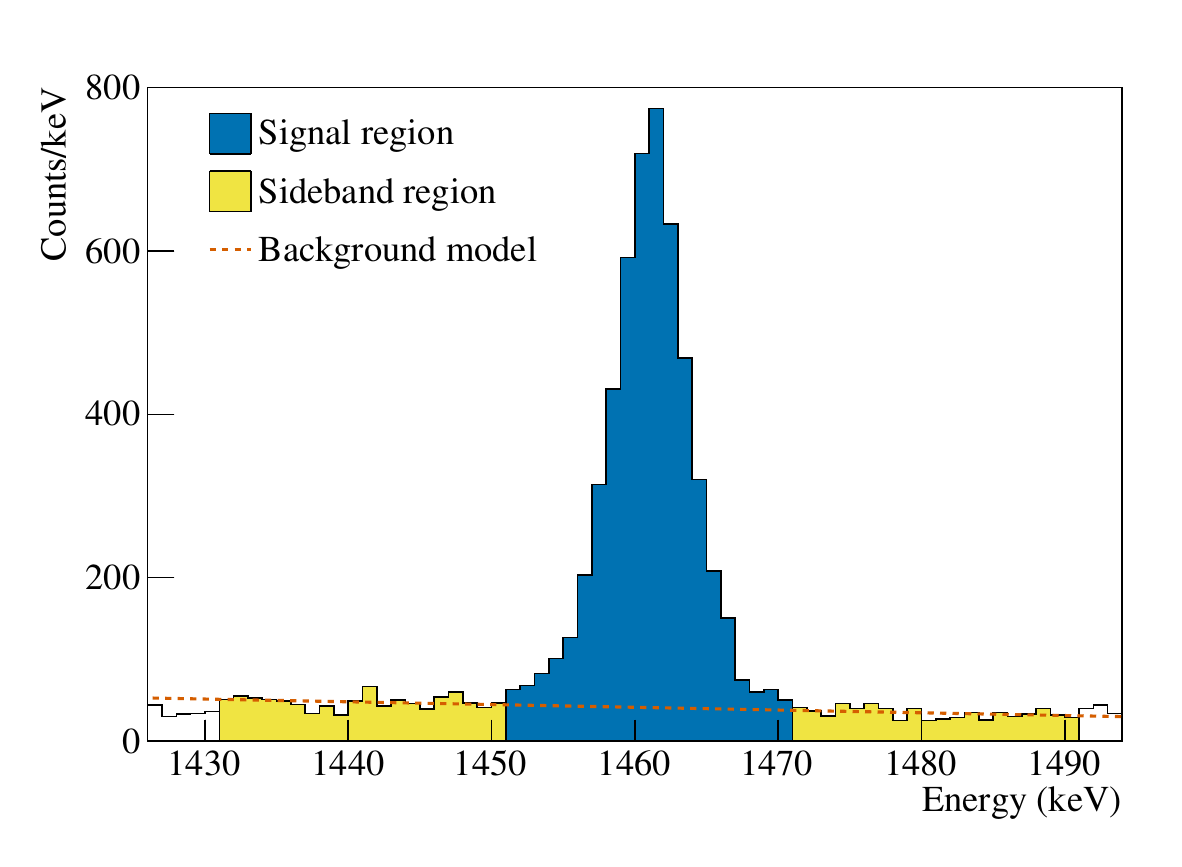}
\caption{Events that pass the basic quality cuts with M\,=\,1 for a single dataset in the $^{40}$K 1461\,keV peak region.} 
\label{fig:fit_pass}     
\end{figure}

\begin{figure}
  \includegraphics[trim={0.5cm 0.6cm 0.9cm 1.2cm},clip,width=\columnwidth]{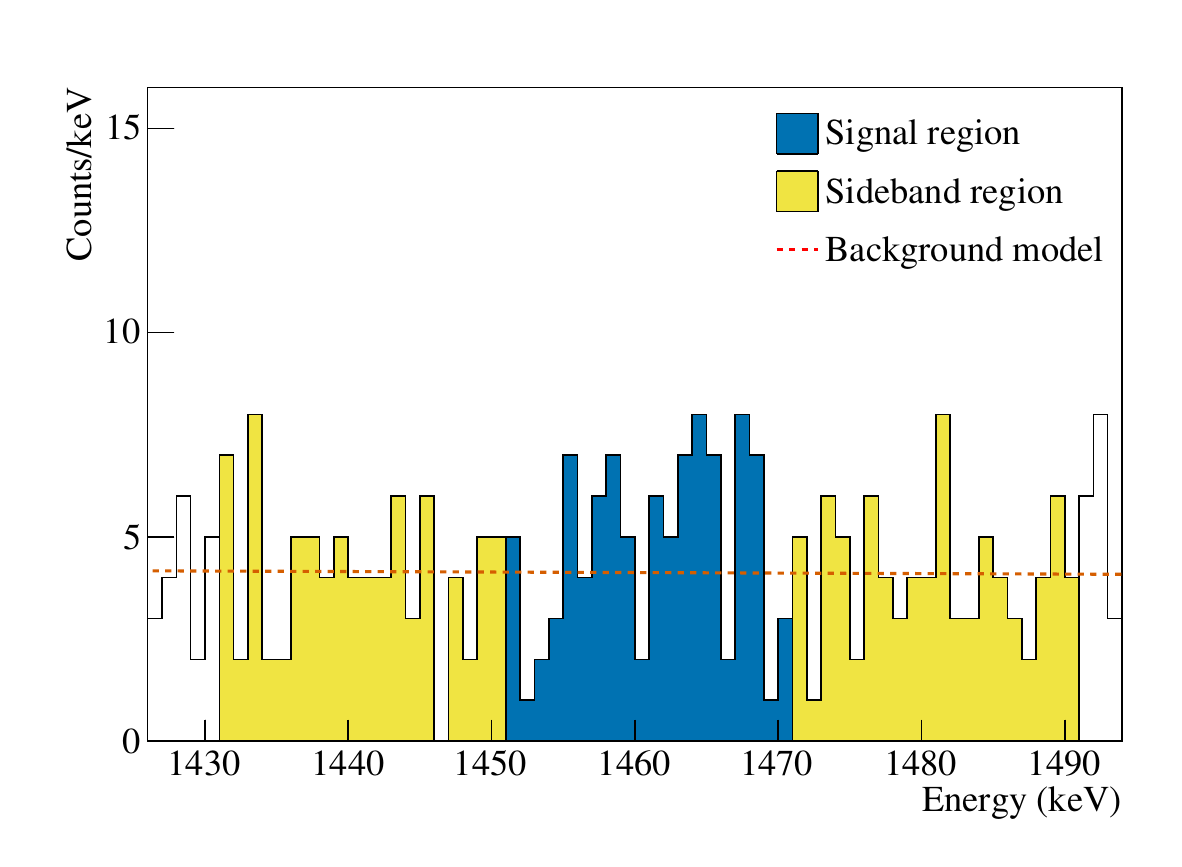}
\caption{Events that pass the basic quality cuts and fail the M\,=\,1 cut for a single dataset in the $^{40}$K 1461\,keV peak region.} 
\label{fig:fit_fail} 
\end{figure}

\begin{figure}
  \includegraphics[trim={0.7cm 0.7cm 0.3cm 0cm},clip,width=\columnwidth]{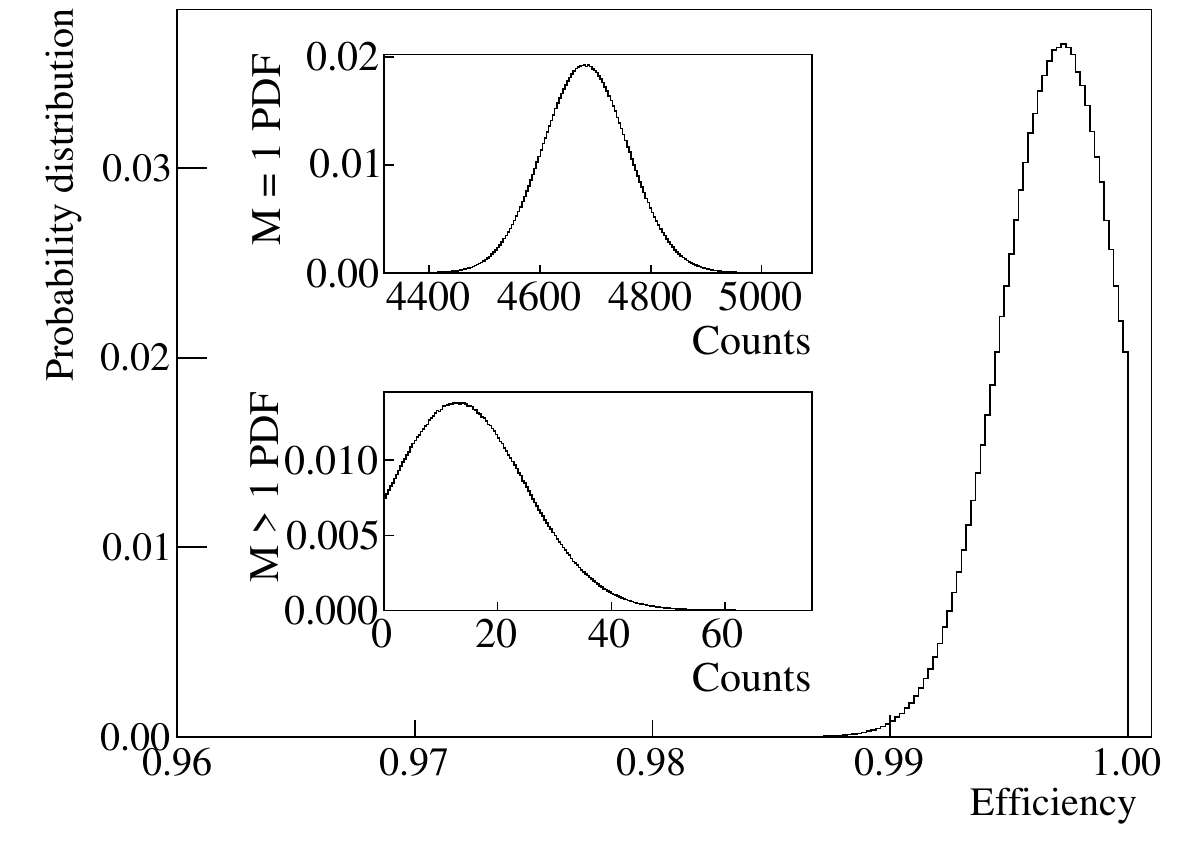}
\caption{AC efficiency prior probability distribution for a single dataset. Posterior for the number of events above background in the $^{40}$K 1461\,keV peak region that passed basic quality cuts with M\,=\,1 (M\,$>$\,1) for the corresponding dataset is shown in the top (bottom) inset.} 
\label{fig:post_eff} 
\end{figure}

The anti-coincidence (AC) efficiency is calculated from the physics data after applying the basic quality cuts. We calculate this dataset-dependent efficiency as
\begin{equation}
\epsilon_{AC}=n_{pass}/(n_{pass}\,+\,n_{fail})
\end{equation}
where $n_{pass}$ and $n_{fail}$ are the number of events above background in the 1461\,keV $^{40}$K $\gamma$ peak signal region that pass and fail the M\,=\,1 cut, respectively. We use the $^{40}$K $\gamma$ peak because it is not associated with a gamma cascade, ensuring that the events in the M\,$>$\,1 spectrum are due to accidental coincidences. 

To evaluate the efficiency, we perform a counting analysis in the $^{40}$K peak region, assuming a Poisson distribution for the number of counts in the signal region and background side bands:  
\begin{equation}
    \centering
    \mathcal{L}_{eff} = \prod_{i=l,c,r} \frac{e^{-\lambda_i}\lambda_i^{n_i}}{n_i!}
\end{equation}
where the index $i$ is over adjacent energy regions---the left side band ($l$), the center peak signal region ($c$), and the right side band ($r$)---and the expected number of counts in each region [$E_{i,1}$, $E_{i,2}$] is
\begin{equation}
    \centering
    \lambda_i = \int_{E_{i,1}}^{E_{i,2}} f(b)dE + S_i
    \label{eq:lambda}
\end{equation}
where $f(b)$ is a linear function describing the background between $E_{l,1}$ and $E_{r,2}$, $S_l=S_r=0$, and $S_c$ is the neat signal count in the peak signal region. We perform a Bayesian counting analysis with BAT to extract the posteriors for the 
neat signal counts. A probability distribution for the AC efficiency is evaluated by taking random samples of the BAT posteriors for $S_{c,pass}$ and $S_{c,fail}$ and combining each sample-pair ($n_{pass}$,$n_{fail}$) to compute $\epsilon_{AC}$. The resulting probability distribution for the AC efficiency on the $^{40}$K $\gamma$ peak is assumed to be the same at $Q_{\beta\beta}$. Figures \ref{fig:fit_pass} and \ref{fig:fit_fail} show the $^{40}$K peak fit region with M\,=\,1 and M\,$>$\,1, respectively, for a single dataset. The corresponding posterior probability distributions for the neat signal counts are shown in Fig.~\ref{fig:post_eff} for M\,=\,1 in the top inset and M\,$>$\,1 in the bottom inset. The AC efficiency probability distribution for the single dataset is shown in Fig.~\ref{fig:post_eff}.

\begin{figure*}
    \centering
\includegraphics[trim={0.3cm 0cm 0.08cm 0cm},clip,width=\textwidth]
{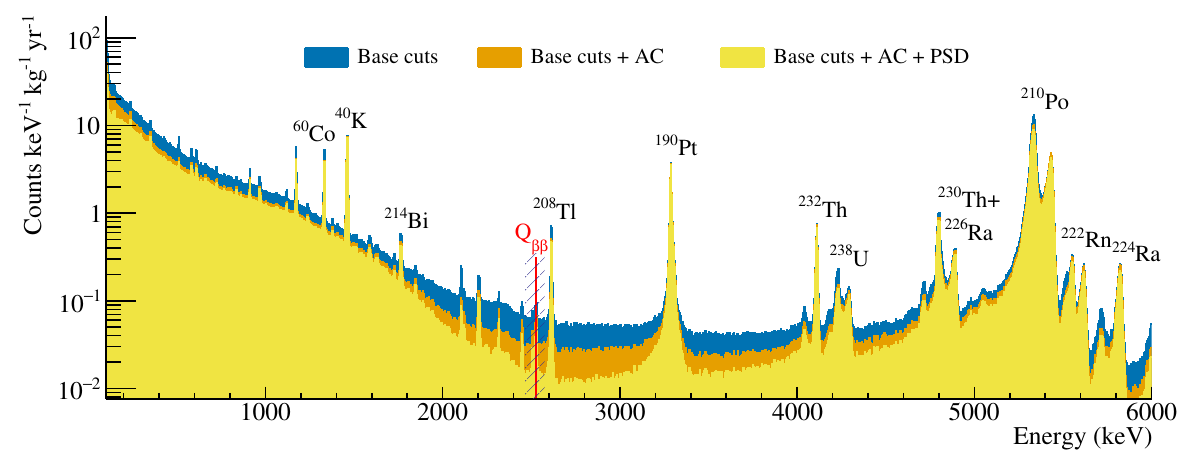}
    \caption{Blinded physics spectrum for the total 2\,\tony data release. The shaded area indicates the salted peak region at Q$_{\beta\beta}$.}
    \label{fig:blinded}
\end{figure*}

\subsection{Pulse shape discrimination efficiency} \label{sec:psdeff}

The pulse shape discrimination (PSD) efficiency is calculated from the physics data after performing basic quality cuts and the AC cut (M\,=\,1). The PSD efficiency is the probability of preserving a physical event upon applying the NRE threshold cut (see Sec.~\ref{sec:psd}). Unlike the AC efficiency, the PSD efficiency is assumed to be energy dependent, so we evaluate the corresponding efficiency on several $\gamma$ peaks in the physics spectrum and extrapolate to $Q_{\beta\beta}$. Only peaks with reasonable fit quality with respect to the lineshape are considered in evaluating the dataset level PSD efficiency, $\epsilon_{PSD}$. Similar to the AC efficiency evaluation, we use a Bayesian counting analysis to extract the posteriors for $S_{c,pass}$ and $S_{c,fail}$ for each of the selected dataset-peaks, then combine their posteriors numerically to obtain the energy-dependent PSD efficiency probability distributions. The probability distribution for $\epsilon_{PSD}$ of each $\gamma$ peak is fit to a Gaussian function in order to obtain a smooth distribution, but only the physical distribution range between 0 and 1 is used. A posterior for the PSD efficiency at $Q_{\beta\beta}$ is extracted by performing a combined fit of the Gaussian distributions to a first-order polynomial function. For the 2\,\tony \onbb decay analysis, the exposure-weighted AC efficiency is 99.80(5)\% and the exposure-weighted PSD efficiency is 97.9(18)\%~\cite{cuore2025:science}.

The total analysis efficiency for each dataset is calculated by combining the PDFs of all the corresponding efficiencies numerically,  
as described in Sec.~\ref{sec:bc}. 
The resulting distribution is used as a dataset-dependent prior PDF for the total efficiency in the \onbb decay fit. The exposure-weighted total analysis efficiency for the \onbb decay analysis is 93.4(18)\%~\cite{cuore2025:science}.

\section{Physics analyses} 

An unprecedented amount of high-quality data has been collected to perform a high-sensitivity search for the \onbb decay of $^{130}$Te. Using the current data analysis procedure, described in this paper, we present the blinded physics spectrum from over 2\,\tony of \teod analyzed exposure in Fig.~\ref{fig:blinded}. After performing our search for the \onbb decay peak in the ROI of the unblinded spectrum, we set the most stringent limit on the \onbb decay half-life of $^{130}$Te. This result is reported in~\cite{cuore2025:science}. 

CUORE leverages the extensive amount of data acquired to undertake a variety of physics analyses. 
One crucial physics analysis involves constructing a detailed background model to identify the radioactive contamination that contributes to our data, as detailed in~\cite{CUORE:2024fak}. This model is essential for improving the sensitivity of our primary search and for the background predictions of the next-generation experiment, CUPID~\cite{CUPID:2025epjc}, which will use the same technology and infrastructure. Our robust background model allows for a precise measurement of the $^{130}$Te 2$\nu\beta\beta$ decay half-life and for dedicated spectral shape studies for the cross-validation of the nuclear models~\cite{cuore2024:buccinu} and searches for Beyond Standard Physics (BSM) effects. In addition, we investigate muon interactions; profiting from the high granularity of the CUORE detector, we reconstruct muon tracks through the detector to further refine our understanding of the background. 
Another significant multi-crystal analysis focuses on the search for Fractionally Charged Particles (FCP)~\cite{CUORE:2024rbd}, which could provide insights to BSM. By employing advanced denoising techniques, we are also able to perform very sensitive low-frequency noise studies on long timescales; this allowed us to identify the impact of marine microseisms on the detector resolution~\cite{Aragao:2024ijn,SeaDen:2024}. Finally, the keV-scale thresholds achieved with the described CUORE data analysis procedure enables low energy studies~\cite{cuore2025:lowE}, which are crucial for investigating many rare event signals, including potential dark matter interactions.

\section{Conclusions}
The CUORE experiment has made significant advances in the development and optimization of analysis tools used to manage and interpret the data collected by its cryogenic calorimeters. These tools, initially built for CUORE prototypes, have been refined to handle the challenges of acquiring and processing data from $\sim$1000 channels. The automated data analysis methods have facilitated the production of high-quality datasets. The optimized analysis tools have allowed CUORE to conduct a high-sensitivity search for the \onbb decay of $^{130}$Te using over 2\,\tony \teod exposure, the largest dataset ever collected and analyzed for this kind of detector technology. Overall, these analysis tools have enhanced our ability to explore rare events, thereby contributing to the overall scientific output of CUORE.

\begin{acknowledgements}
The CUORE Collaboration thanks the directors and staff of the Laboratori Nazionali del Gran Sasso and the technical staff of our laboratories.
This work was supported by the Istituto Nazionale di Fisica Nucleare (INFN); the National Science Foundation under Grant Nos. NSF-PHY-0605119, NSF-PHY-0500337, NSF-PHY-0855314, NSF-PHY-0902171, NSF-PHY-0969852, NSF-PHY-1307204, NSF-PHY-1314881, NSF-PHY-1401832, NSF-PHY-1913374, and NSF-PHY-2412377; Yale University, Johns Hopkins University, and University of Pittsburgh.
This material is also based upon work supported by the US Department of Energy (DOE) Office of Science under Contract Nos. DE-AC02-05CH11231, and DE-AC52-07NA27344; by the DOE Office of Science, Office of Nuclear Physics under Contract Nos. DE-FG02-08ER41551, DE-FG03-00ER41138, DE-SC0012654, DE-SC0\-020423, DE-SC0019316, and DE-SC0011091.
This research used resources of the National Energy Research Scientific Computing Center (NERSC).
This work makes use of both the DIANA data analysis and APOLLO data acquisition software packages, which were developed by the CUORICINO, CUORE, LUCIFER, and CUPID-0 Collaborations.
The authors acknowledge the Advanced Research Computing at Virginia Tech and the Yale Center for Research Computing for providing computational resources and technical support that have contributed to the results reported within this paper.
\end{acknowledgements}

\bibliographystyle{spphys}       
\bibliography{bibliography}   

\begin{thebibliography}{10}
\providecommand{\url}[1]{{#1}}
\providecommand{\urlprefix}{URL }
\expandafter\ifx\csname urlstyle\endcsname\relax
  \providecommand{\doi}[1]{DOI \discretionary{}{}{}#1}\else
  \providecommand{\doi}{DOI \discretionary{}{}{}\begingroup
  \urlstyle{rm}\Url}\fi

\bibitem{cuore2025:science}
D.Q. Adams, et~al., Science \textbf{0}(0), eadp6474 (2025).
\newblock \doi{10.1126/science.adp6474}

\bibitem{CUORE:2021mvw}
D.Q. Adams, et~al., Nature \textbf{604}(7904), 53 (2022).
\newblock \doi{10.1038/s41586-022-04497-4}

\bibitem{CUPID:2022puj}
O.~Azzolini, et~al., Phys. Rev. Lett. \textbf{129}(11), 111801 (2022).
\newblock \doi{10.1103/PhysRevLett.129.111801}

\bibitem{Agrawal:2025}
A.~Agrawal, et~al., Phys. Rev. Lett. \textbf{134}, 082501 (2025).
\newblock \doi{10.1103/PhysRevLett.134.082501}

\bibitem{GERDA:PRL}
M.~Agostini, et~al., Phys. Rev. Lett. \textbf{125}, 252502 (2020).
\newblock \doi{10.1103/PhysRevLett.125.252502}

\bibitem{KamLAND-Zen:2022tow}
S.~Abe, et~al., Phys. Rev. Lett. \textbf{130}(5), 051801 (2023).
\newblock \doi{10.1103/PhysRevLett.130.051801}

\bibitem{FrontendCUORE}
C.~Arnaboldi, et~al., JINST \textbf{13}(02), P02026 (2018).
\newblock \doi{10.1088/1748-0221/13/02/P02026}

\bibitem{LinearSupplyCUORE}
P.~Carniti, et~al., Review of Scientific Instruments \textbf{87}(5), 054706
  (2016).
\newblock \doi{10.1063/1.4948390}

\bibitem{PulserCUORE}
K.~Alfonso, et~al., JINST \textbf{13}(02), P02029 (2018).
\newblock \doi{10.1088/1748-0221/13/02/P02029}

\bibitem{CUORE:2022hzh}
D.Q. Adams, et~al., JINST \textbf{17}(11), P11023 (2022).
\newblock \doi{10.1088/1748-0221/17/11/P11023}

\bibitem{Alfonso:2020yee}
K.~Alfonso, et~al., Nucl. Instrum. Methods A \textbf{1008}, 165451 (2021).
\newblock \doi{10.1016/j.nima.2021.165451}

\bibitem{Dompe:2020JLTP}
V.~{Domp{\`e}}, et~al., J. Low Temp. Phys. \textbf{200}(5-6), 286 (2020).
\newblock \doi{10.1007/s10909-020-02435-0}

\bibitem{DAddabbo:2017efe}
A.~D'Addabbo, et~al., Cryogenics \textbf{93}, 56 (2018).
\newblock \doi{10.1016/j.cryogenics.2018.05.001}

\bibitem{Alduino:2016vjd}
C.~Alduino, et~al., JINST \textbf{11}(07), P07009 (2016).
\newblock \doi{10.1088/1748-0221/11/07/P07009}

\bibitem{BIASSONI2020103803}
M.~Biassoni, O.~Cremonesi, Progress in Particle and Nuclear Physics
  \textbf{114}, 103803 (2020).
\newblock \doi{10.1016/j.ppnp.2020.103803}

\bibitem{Rahaman:2011zz}
S.~Rahaman, et~al., Phys. Lett. B \textbf{703}, 412 (2011).
\newblock \doi{10.1016/j.physletb.2011.07.078}

\bibitem{ANDREOTTI2012161}
E.~Andreotti, et~al., Nucl. Instrum. Methods A \textbf{664}(1), 161  (2012).
\newblock \doi{10.1016/j.nima.2011.10.065}

\bibitem{Nutini:2020vtd}
I.~Nutini, J. Low Temp. Phys. \textbf{199}(1-2), 519 (2020).
\newblock \doi{10.1007/s10909-020-02402-9}

\bibitem{ALDUINO:cryo2019}
C.~Alduino, et~al., Cryogenics \textbf{102}, 9  (2019).
\newblock \doi{10.1016/j.cryogenics.2019.06.011}

\bibitem{CUORE:2021ctv}
D.Q. Adams, et~al., Progress in Particle and Nuclear Physics \textbf{122},
  103902 (2021).
\newblock \doi{10.1016/j.ppnp.2021.103902}

\bibitem{cuore2024:buccinu}
C.~Bucci, et~al.
\newblock {CUORE latest results and prospects} (2024).
\newblock \doi{10.5281/zenodo.12706030}.
\newblock Talk given at the Neutrino 2024 Conference, Milano, Italy, June, 2024

\bibitem{Alduino:2017ehq}
C.~Alduino, et~al., Phys. Rev. Lett. \textbf{120}(13), 132501 (2018).
\newblock \doi{10.1103/PhysRevLett.120.132501}

\bibitem{Adams:2019jhp}
D.~Adams, et~al., Phys. Rev. Lett. \textbf{124}(12), 122501 (2020).
\newblock \doi{10.1103/PhysRevLett.124.122501}

\bibitem{DiDomizio:2018ldc}
S.~Di~Domizio, et~al., JINST \textbf{13}(12), P12003 (2018).
\newblock \doi{10.1088/1748-0221/13/12/P12003}

\bibitem{BRUN199781}
R.~Brun, F.~Rademakers, Nucl. Instrum. Methods A \textbf{389}(1), 81 (1997).
\newblock \doi{10.1016/S0168-9002(97)00048-X}

\bibitem{Alduino:2016zrl}
C.~Alduino, et~al., Phys. Rev. C \textbf{93}(4), 045503 (2016).
\newblock \doi{10.1103/PhysRevC.93.045503}

\bibitem{vetter2024improving}
K.J. Vetter, et~al., Eur. Phys. J. C \textbf{84}(3), 243 (2024).
\newblock \doi{10.1140/epjc/s10052-024-12595-y}

\bibitem{Bucci_2017}
C.~Bucci, et~al., JINST \textbf{12}(12), P12013 (2017).
\newblock \doi{10.1088/1748-0221/12/12/P12013}

\bibitem{Alduino:2017xpk}
C.~Alduino, et~al., Eur. Phys. J. C \textbf{77}(12), 857 (2017).
\newblock \doi{10.1140/epjc/s10052-017-5433-1}

\bibitem{Branca:2020bra}
A.~Branca, J. Phys. Conf. Ser. \textbf{1468}, 012118 (2020).
\newblock \doi{10.1088/1742-6596/1468/1/012118}

\bibitem{Gatti:1986cw}
E.~Gatti, P.F. Manfredi, Riv. Nuovo Cim. \textbf{9N1}, 1 (1986).
\newblock \doi{10.1007/BF02822156}

\bibitem{ANDREOTTI2011822}
E.~Andreotti, et~al., Astrop. Phys. \textbf{34}(11), 822 (2011).
\newblock \doi{10.1016/j.astropartphys.2011.02.002}

\bibitem{Alessandrello:1998}
A.~Alessandrello, et~al., Nucl. Instrum. Methods A \textbf{412}(2), 454
  (1998).
\newblock \doi{10.1016/S0168-9002(98)00458-6}

\bibitem{cuore2025:lowE}
D.Q. Adams, et~al.,   (2025).
\newblock Paper in preparation.

\bibitem{CUORE:2024rbd}
D.Q. Adams, et~al., Phys. Rev. Lett. \textbf{133}, 241801 (2024).
\newblock \doi{10.1103/PhysRevLett.133.241801}

\bibitem{Alduino_2017}
C.~Alduino, et~al., Eur. Phys. J. C \textbf{77}(8) (2017).
\newblock \doi{10.1140/epjc/s10052-017-5080-6}

\bibitem{Huang_PhD-thesis:2021}
R.G. Huang,   (2021).
\newblock [Ph.\,D.\ thesis, University of California - Berkeley]

\bibitem{Adams:2022hji}
D.Q. Adams, et~al., Phys. Rev. Lett. \textbf{129}, 222501 (2022).
\newblock \doi{10.1103/PhysRevLett.129.222501}

\bibitem{CUORE:2022dwv}
D.Q. Adams, et~al., Phys. Rev. C \textbf{105}, 065504 (2022).
\newblock \doi{10.1103/PhysRevC.105.065504}

\bibitem{BATSoftware}
A.~Caldwell, D.~Koll\'{a}r, K.~Kr\"{o}ninger, Computer Physics Communications
  \textbf{180}, 2197 (2009).
\newblock \doi{10.1016/j.cpc.2009.06.026}

\bibitem{CUORE:2024fak}
D.Q. Adams, et~al., Phys. Rev. D \textbf{110}, 052003 (2024).
\newblock \doi{10.1103/PhysRevD.110.052003}

\bibitem{CUPID:2025epjc}
K.~Alfonso, et~al., Eur. Phys. J. C \textbf{85}, 737 (2025).
\newblock \doi{10.1140/epjc/s10052-025-14352-1}

\bibitem{Aragao:2024ijn}
L.~Arag\~ao, et~al., Eur. Phys. J. C \textbf{84}(7), 728 (2024).
\newblock \doi{10.1140/epjc/s10052-024-13065-1}

\bibitem{SeaDen:2024}
D.Q. Adams, et~al.,   (2025).
\newblock Paper in preparation.

\end{thebibliography}

\end{document}